\documentclass{article}

\PassOptionsToPackage{numbers, sort&compress}{natbib}

\usepackage[preprint]{neurips_2024}

\usepackage[utf8]{inputenc} 
\usepackage[T1]{fontenc}    
\usepackage{hyperref}       
\usepackage{url}            
\usepackage{booktabs}       
\usepackage{amsfonts}       
\usepackage{nicefrac}       
\usepackage{microtype}      
\usepackage{xcolor}         

\usepackage{wrapfig}
\usepackage{graphicx} 		
\usepackage{tikz}

\def\astrolink{\texttt{AstroLink}}
\def\fuzzycat{\texttt{FuzzyCat}}

\newcommand{\hMpc}{{\ifmmode{h^{-1}{\rm Mpc}}\else{$h^{-1}$Mpc}\fi}}
\newcommand{\Mpc}{{\ifmmode{{\rm Mpc}}\else{Mpc}\fi}}
\newcommand{\hkpc}{{\ifmmode{h^{-1}{\rm kpc}}\else{$h^{-1}$kpc}\fi}}
\newcommand{\kpc}{{\ifmmode{ {\rm kpc} }\else{{\rm kpc}}\fi}}
\newcommand{\kms}{{\ifmmode{ {\rm km\,s^{-1}} }\else{ ${\rm km\,s^{-1}}$ }\fi}}
\newcommand{\hMsun}{{\ifmmode{h^{-1}{\rm {M_{\odot}}}}\else{$h^{-1}{\rm{M_{\odot}}}$}\fi}}
\newcommand{\Msun}{{\ifmmode{{\rm M}_{\odot}}\else{${\rm M}_{\odot}$}\fi}}
\newcommand{\Mhalo}{{\ifmmode{M_{\rm halo}}\else{$M_{\rm halo}$}\fi}}
\newcommand{\Rvir}{{\ifmmode{R_{\rm vir}}\else{$R_{\rm vir}$}\fi}}
\newcommand{\Mvir}{{\ifmmode{M_{\rm vir}}\else{$M_{\rm vir}$}\fi}}
\newcommand{\Mstar}{{\ifmmode{M_{\rm star}}\else{$M_{\rm star}$}\fi}}
\newcommand{\Vrot}{{\ifmmode{V_{\rm rot}}\else{$V_{\rm rot}$}\fi}}
\newcommand{\ltsima}{$\; \buildrel < \over \sim \;$}
\newcommand{\gtsima}{$\; \buildrel > \over \sim \;$}
\newcommand{\lsim}{\lower.5ex\hbox{\ltsima}}
\newcommand{\gsim}{\lower.5ex\hbox{\gtsima}}

\def\lesssim{\mathrel{\hbox{\rlap{\hbox{\lower4pt\hbox{$\sim$}}}\hbox{$<$}}}}
\def\gtrsim{\mathrel{\hbox{\rlap{\hbox{\lower4pt\hbox{$\sim$}}}\hbox{$>$}}}}

\newcommand{\beq}{\begin{equation}}
\newcommand{\eeq}{\end{equation}}
\def\beqa{\begin{eqnarray}}
\def\eeqa{\end{eqnarray}}
\def\LCDM{\ensuremath{\Lambda}CDM}

\def \kms {\ifmmode  \,\rm km\,s^{-1} \else $\,\rm km\,s^{-1}  $ \fi }
\def \kpc {\ifmmode  {\rm kpc}  \else ${\rm  kpc}$ \fi  }  
\def \hkpc {\ifmmode  {h^{-1}\rm kpc}  \else ${h^{-1}\rm kpc}$ \fi  }  
\def \hMpc {\ifmmode  {h^{-1}\rm Mpc}  \else ${h^{-1}\rm Mpc}$ \fi  }  
\def \Mpch {\ifmmode  {h^{-1}\rm Mpc}  \else ${h^{-1}\rm Mpc}$ \fi  }  
\def \Msun {\ifmmode {\rm M}_{\odot} \else ${\rm M}_{\odot}$ \fi} 
\def \hMsun {\ifmmode h^{-1}\,\rm M_{\odot} \else $h^{-1}\,\rm M_{\odot}$ \fi}

\def \LCDM {\ifmmode \Lambda{\rm CDM} \else $\Lambda{\rm CDM}$ \fi}
\def \sig8 {\ifmmode \sigma_8 \else $\sigma_8$ \fi} 
\def \OmegaM {\ifmmode \Omega_{\rm m} \else $\Omega_{\rm m}$ \fi} 
\def \Omegab {\ifmmode \Omega_{\rm b} \else $\Omega_{\rm b}$ \fi} 
\def \OmegaL {\ifmmode \Omega_{\rm \Lambda} \else $\Omega_{\rm \Lambda}$\fi} 
\def \Deltavir {\ifmmode \Delta_{\rm vir} \else $\Delta_{\rm vir}$ \fi}
\def \rhocrit {\ifmmode \rho_{\rm crit} \else $\rho_{\rm crit}$ \fi}
\def \rhou {\ifmmode \rho_{\rm u} \else $\rho_{\rm u}$ \fi}
\def \zc {\ifmmode z_{\rm c} \else $z_{\rm c}$ \fi}

\title{Galaxy Formation and Evolution via Phase-temporal Clustering with FuzzyCat $\circ$ AstroLink}

\author{%
  William H. Oliver {\rm and} Tobias Buck \\
  Interdisciplinary Center for Scientific Computing, University of Heidelberg \\
  Im Neuenheimer Feld 205, D-69120 Heidelberg, Germany \\
  \href{mailto:william.oliver@iwr.uni-heidelberg.de}{\texttt{william.oliver@iwr.uni-heidelberg.de}} \& \href{mailto:tobias.buck@iwr.uni-heidelberg.de}{\texttt{tobias.buck@iwr.uni-heidelberg.de}}
}

\begin{document}

\maketitle

\begin{abstract}
We demonstrate how the composition of two unsupervised clustering algorithms, \astrolink\ and \fuzzycat, makes for a powerful tool when studying galaxy formation and evolution. \astrolink\ is a general-purpose astrophysical clustering algorithm built for extracting meaningful hierarchical structure from point-cloud data defined over any feature space, while \fuzzycat\ is a generalised soft-clustering algorithm that propagates the dynamical effects of underlying data processes into a fuzzy hierarchy of stable fuzzy clusters.
Their composition, \fuzzycat\ $\circ$ \astrolink, can therefore identify a fuzzy hierarchy of astrophysically- and statistically-significant fuzzy clusters within any point-based data set whose representation is subject to changes caused by some underlying process. Furthermore, the pipeline achieves this without relying upon strong assumptions about the data, the change process, the number/importance of specific structure types, or much user input -- thereby making itself applicable to a wide range of fields in the physical sciences.
We find that for the task of structurally decomposing simulated galaxies into their constituents, our context-agnostic approach has a substantial impact on the diversity and completeness of the structures extracted as well as on their relationship within the broader galactic structural hierarchy -- revealing dwarf galaxies, infalling groups, stellar streams (and their progenitors), stellar shells, galactic bulges, and star-forming regions.
\end{abstract}

\section{Motivation and related work}

A pressing and continually evolving sub-field of astrophysics is the study of galaxy formation and evolution, which according to $\Lambda$CDM cosmology, assemble hierarchically with time through mergers and the accretion of smaller galaxies \citep{Springel2005b}. To understand how and why a galaxy and its substructure develop within the context of the surrounding environment and of the underlying cosmological model, astrophysicists and cosmologists will look to both observational and simulation data. With observations, we may learn from a very large number of galaxies that are each observed at a unique snapshot in time and that arise from the \textit{ground-truth} cosmology of our Universe. While with simulations, we may learn from many snapshots of a comparatively small number of galaxies that depend on a pre-specified cosmological model. By comparing these two data types, we can hope to constrain our cosmological models as well as our understanding of galaxy formation and evolution.

A typical approach towards studying galaxy formation and evolution in the context of simulated data is to use a halo finder (+ merger tree) code \citep[e.g.][]{Springel2001, Knollmann2009, Behroozi2012, Han2018} to find a catalogue of (sub-)haloes and their merger history (and then to analyse the physical properties of their outputs). These codes routinely perform similarly and robustly \citep{Knebe2011, Knebe2013a, Onions2012, Onions2013, Elahi2013, Avila2014, Lee2014, Behroozi2015}, however they only consider mostly or completely self-bound groups that satisfy a minimum overdensity threshold. If this threshold is too high then some haloes may not be detected and if it is too low then some haloes can be disregarded in the unbinding procedure. As such, these codes will tend not to capture tidally disrupted groups or fleeting structures resulting from density waves, hydrodynamical effects, or star-formation events. Not including these kinds of structures in any subsequent analysis ensures that cosmological models used in simulations are never constrained against their existence -- even though their analogues are observed to be present in our Universe \citep{McConnachie2018, Malhan2022, MiroCarretero2024}. It is for these reasons that we investigate the usefulness of a more generalised and context-agnostic approach.

\section{Finding stable clusters from evolving data: FuzzyCat $\circ$ AstroLink}

The \fuzzycat\ $\circ$ \astrolink\ pipeline operates on a data set whose representation is subject to changes from an underlying process (such as stochastic resampling, temporal-evolution, etc.). The approach composes the two clustering algorithms by first applying \astrolink\ \citep{Oliver2024, AstroLinkGithub2024} to the various realisations of the data and then applying \fuzzycat\ \citep{FuzzyCatGithub2024} to the various \astrolink\ outputs. The result is an unsupervised machine learning pipeline that produces a fuzzy hierarchy of astrophysically- and statistically-significant fuzzy clusters that encapsulate the effects of the underlying process(es) implicit within an evolving input data set. To our knowledge, such a pipeline has not yet existed.

\subsection{The AstroLink algorithm}

\astrolink\ is an unsupervised astrophysical clustering algorithm that extracts arbitrarily-shaped hierarchical clusters from an arbitrarily-shaped point-based data set such that the clusters found are statistical outliers from noisy density fluctuations. It is an improvement to its predecessors, \texttt{CluSTAR-ND} \citep{Oliver2022} and \texttt{Halo-OPTICS} \citep{Oliver2020}, and by comparison boasts increased clustering power in shorter run-times. It also shares algorithmic ties to, but is more statistically robust than, \texttt{OPTICS} \citep{Ankerst1999} and \texttt{HDBSCAN} \citep{Campello2015, McInnes2017}. These can be thought of hierarchical extensions of \texttt{DBSCAN} \citep{Ester1996}, which itself can be thought of as a more-robust-to-noise version of the Friends-Of-Friends (\texttt{FOF}) algorithm \citep{Davis1985} -- an algorithm commonly used to identify galaxies/haloes from cosmological simulations.

The \astrolink\ algorithm performs five steps; (1) data rescaling, (2) local-density estimation, (3) data aggregation, (4) model-fitting and structure identification, (5) hierarchy correction. If \texttt{adaptive} $=1$ (default), step 1 rescales the data to have unit variance -- so as to remove the effect of differing units in the feature space. Step 2 calculates the local-density of each data point by applying a multivariate Epanechnikov kernel \citep{Epanechnikov1969} and a balloon estimator \citep{Sain2002} to its $k_\mathrm{den}$-neighbourhood ($k_\mathrm{den}=20$, default) -- the logarithm of this estimate is taken before all values are then normalised, i.e. $\log\hat\rho \in [0, 1]$. Step 3 tracks and records the connected components of data points that form as the edges of a local-density-weighted $k_{\mathrm{link}}$-nearest-neighbour graph ($k_\mathrm{link}$ is data-driven by default) are traversed in descending order -- these components define a hierarchy of feature-space overdensities. In step 4, a model is fit to the \textit{clusteredness} of these connected components and is used to identify the $\geq S\sigma$-outlier overdensities ($S$ is data-driven by default) from the noisy local-density fluctuations inherent within the data. If \texttt{h\_style} $=1$ (default), step 5 corrects the final hierarchy by incorporating some additional outlier overdensities -- producing the final hierarchy of clusters. \astrolink\ does not require the user to make any hyperparameter choices as the performance of the entirely data-driven version of this process is near-optimal in nearly all cases. When applied to simulated galaxies, \astrolink\ does extraordinarily well at finding the remnants of infalling-satellites within the data. The implementation is described in more detail in the original science paper \citep{Oliver2024} as well as in the \astrolink\ ReadTheDocs page \citep{AstroLinkReadTheDocs2024}.

\subsection{The FuzzyCat algorithm}

\fuzzycat\ is an unsupervised general-purpose soft-clustering algorithm that, given a series of clusterings on object-based data, produces data-driven fuzzy clusters whose membership functions encapsulate the effects of changes in the clusters due to changes in the feature space representation of the objects themselves. The different input clusterings may be governed by any underlying process that affects the clustering structure (e.g. stochasticity, temporal evolution, model hyperparameter variation, etc.). In effect, \fuzzycat\ propagates these effects into a soft-clustering which has had these effects abstracted away into the membership functions of the original object-based data.

At its core, \fuzzycat\ is very similar to \astrolink\ -- procedures mimicking steps 2, 3, and 4 are analogously performed -- except that it takes a data set of clusters as input as opposed to one of data points. It is in this sense that \fuzzycat\ is actually a clustering algorithm that operates on the Jaccard-space of a catalogue of clusterings in order to produce \textit{clusters} of clusters (i.e. fuzzy clusters). As such the Jaccard index, which is calculated for every pair of clusters in the input, is analogous to the \astrolink\ $\log\hat\rho$ calculated in step 2 -- although in \fuzzycat\ there is no equivalent to $k_\mathrm{den}$ and $k_\mathrm{link}$ is effectively fixed at the size of the data set. The final fuzzy clusters, found after a process equivalent to the \astrolink\ step 3, must also meet thresholds (measured by the Jaccard index) of internal similarity ($J_\mathrm{min\_intra} = 0.5$, default) and external dissimilarity ($J_\mathrm{max\_inter} = 0.5$, default), as well as remain stable over at least a minimum number of data set realisations. The latter condition is governed by the \texttt{minStability} hyperparameter, which we change from the default value of $0.5$ for the applications in Sec. \ref{sec:phasetemporalclustering}. These conditions ensure statistical robustness with the corresponding hyperparameters effectively playing the role of the \astrolink\ $S$-parameter. The final fuzzy clusters are then translated into membership functions with respect to the underlying object-based data by counting the number of data realisations for which each object appears within each fuzzy cluster.

\begin{figure}
\vspace{-0.14cm}
\begin{center}
\begin{tikzpicture}
    \vspace{-0.65cm}
    %
    %
    \node (img1)
    {\includegraphics[width=0.2667\textwidth,trim={5mm 15mm 0mm 10mm},clip]{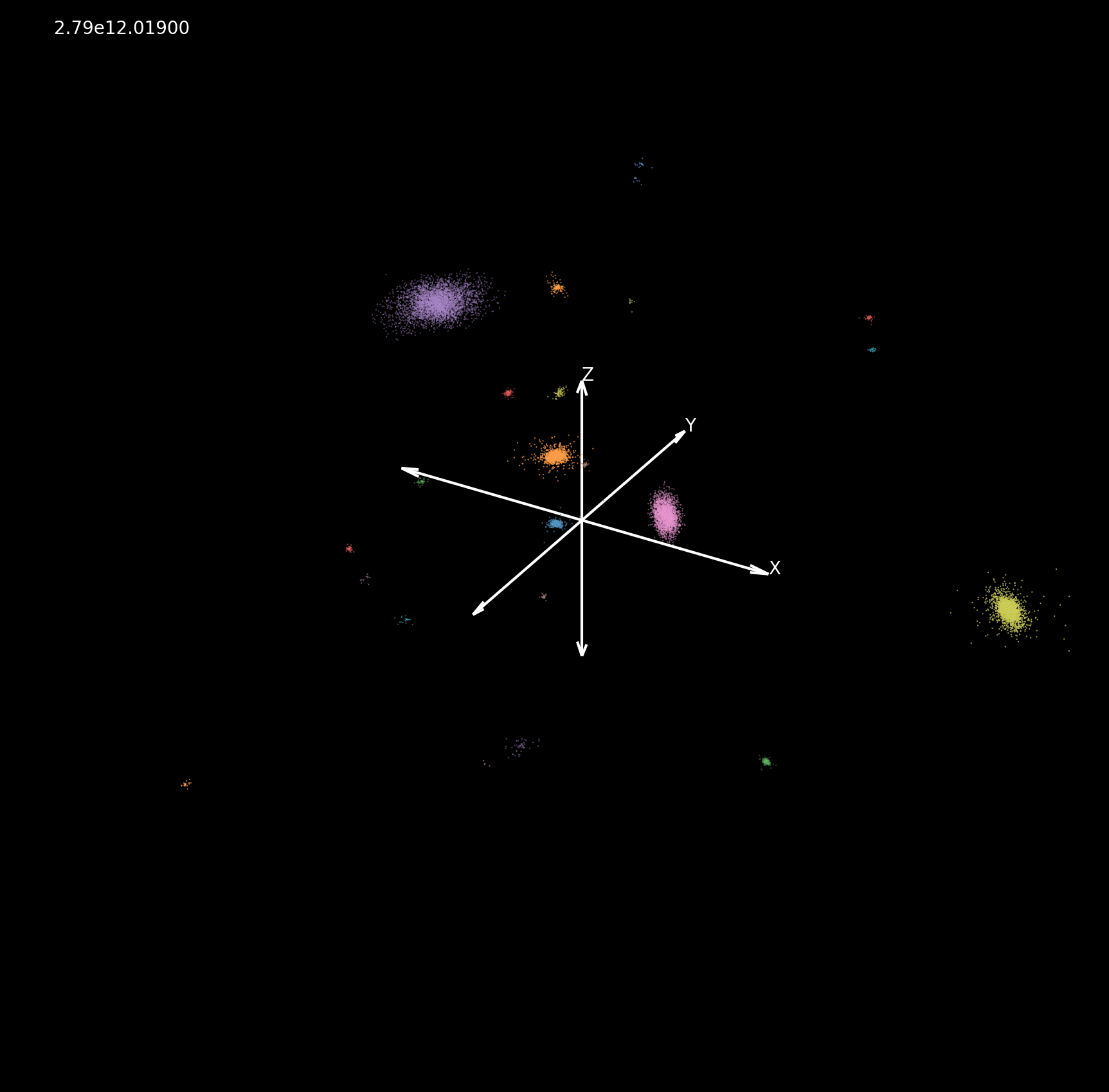}};
    \node at (img1.north) [yshift = 0.013\textwidth] {\large \textbf{AHF}};
    \node at (img1.west) [xshift = -0.075\textwidth, align=center, text height = 1cm] {2.79e12.01900\\$\sim 690$ Myr};
    \node (txt0) at (img1.north west) [xshift = 0.105\textwidth, yshift = -0.03\textwidth] {\footnotesize \textcolor{white}{(sub)haloes}};
    \draw[->, white, thick] (txt0) -- +(-0.055\textwidth, -0.155\textwidth);
    \draw[->, white, thick] (txt0) -- +(0.005\textwidth, -0.035\textwidth);
    \draw[->, white, thick] (txt0) -- +(0.0325\textwidth, -0.035\textwidth);
    \draw[->, white, thick] (txt0) -- +(0.14\textwidth, -0.11\textwidth);
    \node (img2) at (img1.south) [yshift = -0.114\textwidth] 
    {\includegraphics[width=0.2667\textwidth,trim={5mm 15mm 0mm 10mm},clip]{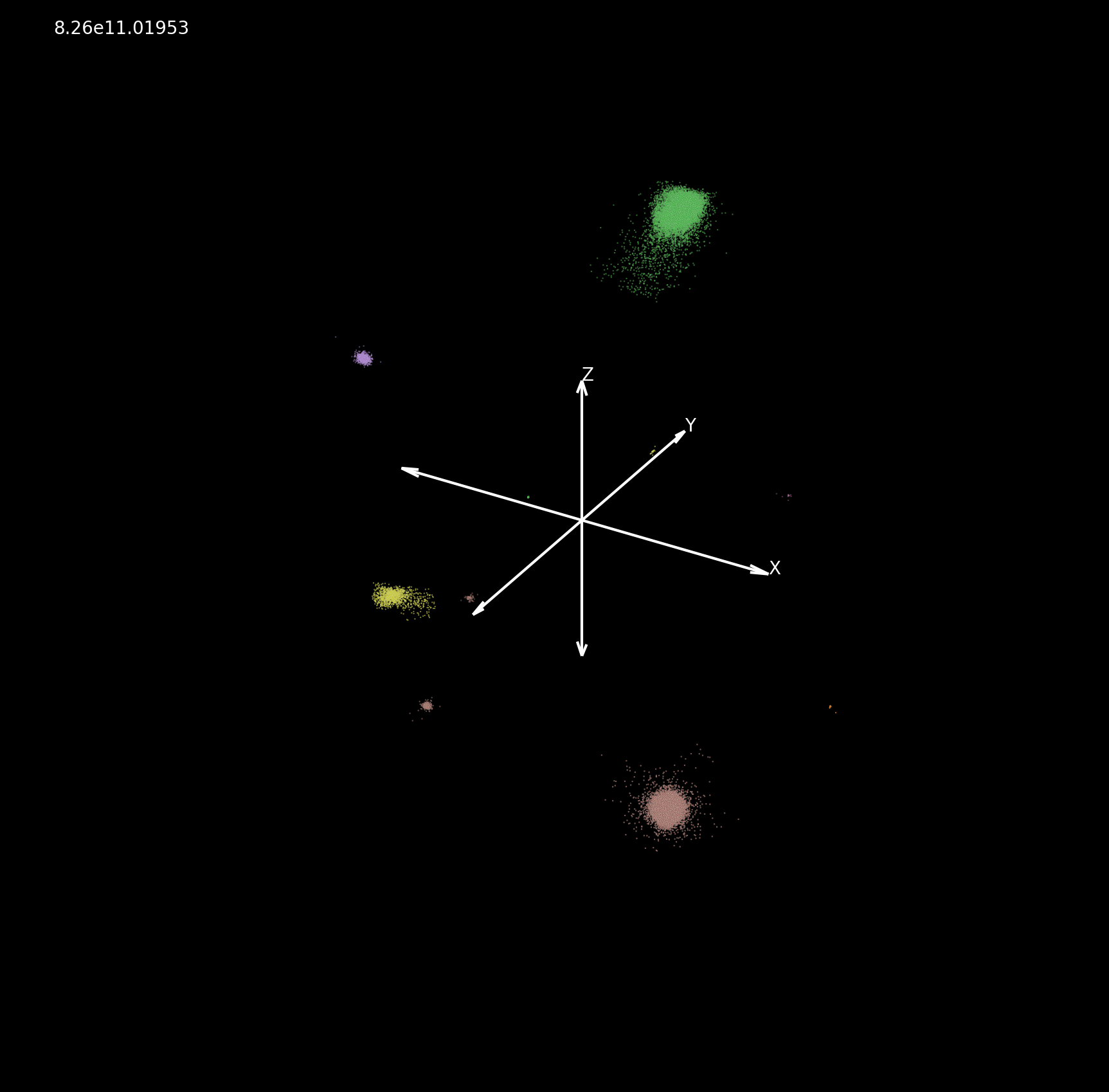}};
    \node at (img2.west) [xshift = -0.075\textwidth, align=center] {8.26e11.01953\\$\sim 324$ Myr};
    \node (img3) at (img2.south) [yshift = -0.114\textwidth]
    {\includegraphics[width=0.2667\textwidth,trim={5mm 15mm 0mm 10mm},clip]{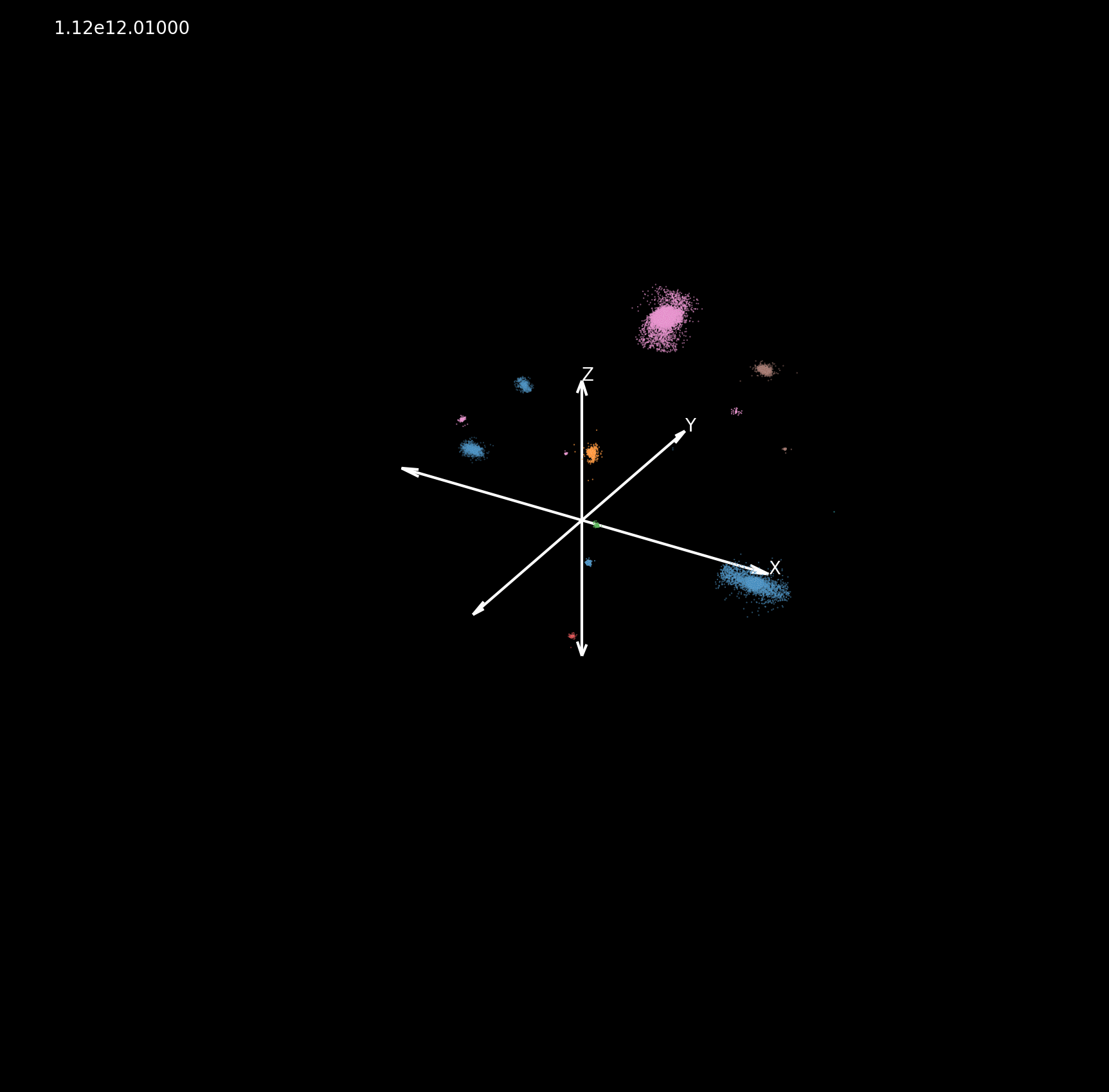}};
    \node at (img3.west) [xshift = -0.075\textwidth, align=center] {1.12e12.01000\\$\sim 6.9$ Gyr};
    \node (img4) at (img3.south) [yshift = -0.114\textwidth]
    {\includegraphics[width=0.2667\textwidth,trim={5mm 15mm 0mm 10mm},clip]{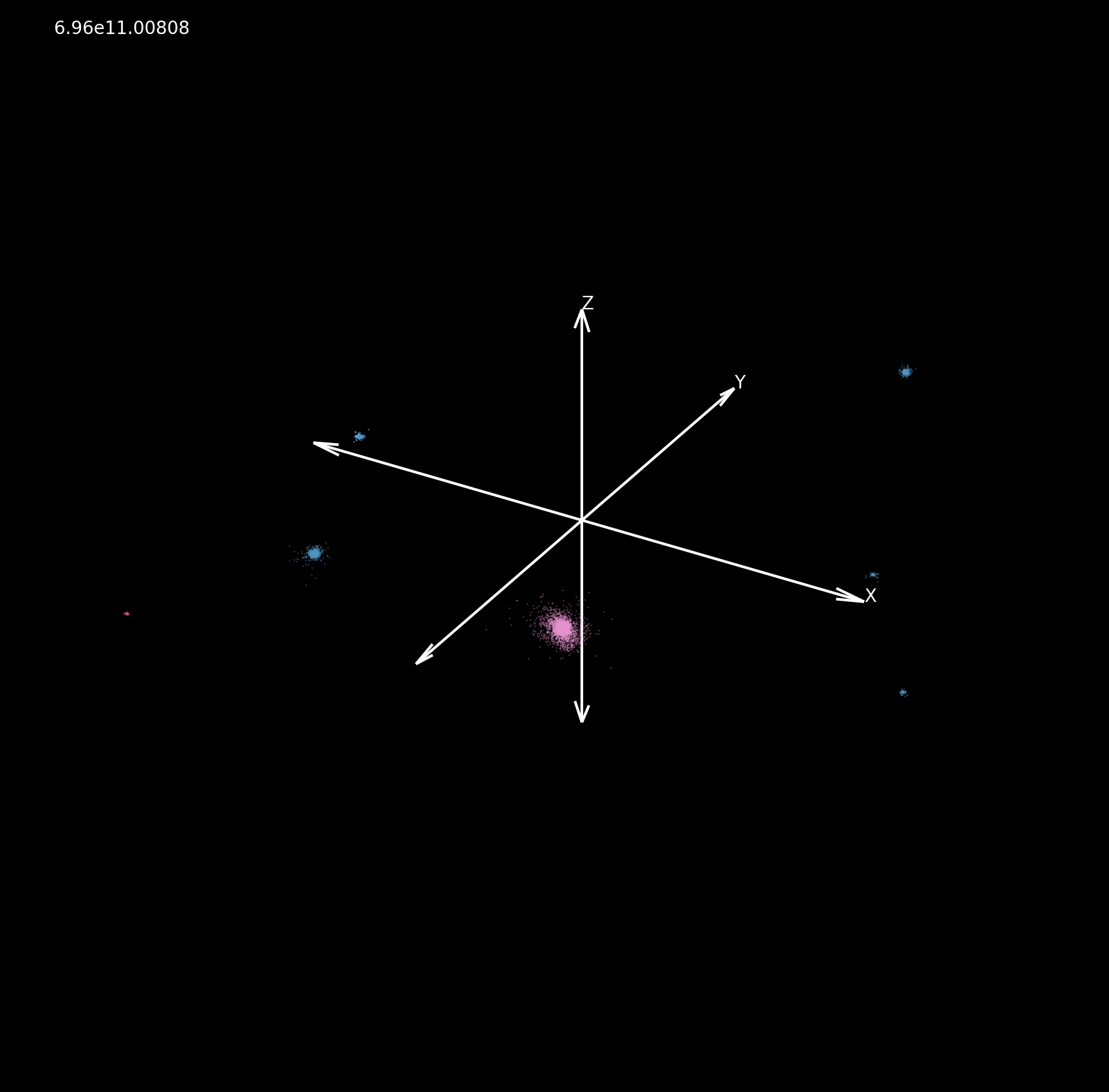}};
    \node at (img4.west) [xshift = -0.075\textwidth, align=center] {6.96e11.00808\\$\sim 2.9$ Gyr};
    \node (img5) at (img4.south) [yshift = -0.114\textwidth]
    {\includegraphics[width=0.2667\textwidth,trim={5mm 15mm 0mm 10mm},clip]{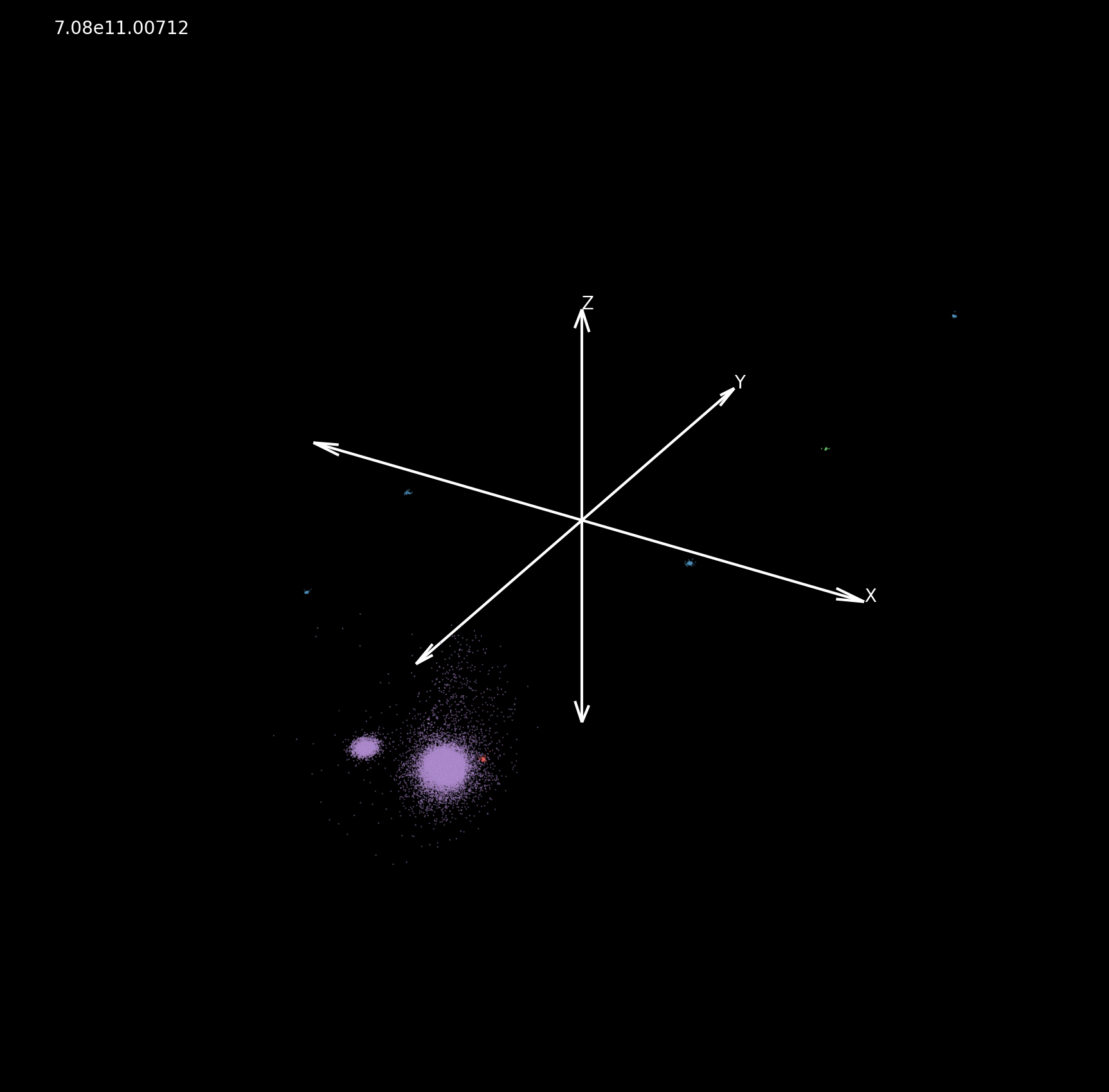}};
    \node at (img5.west) [xshift = -0.075\textwidth, align=center] {7.08e11.00712\\$\sim 4.2$ Gyr};
    \node (img6) at (img5.south) [yshift = -0.114\textwidth]
    {\includegraphics[width=0.2667\textwidth,trim={5mm 15mm 0mm 10mm},clip]{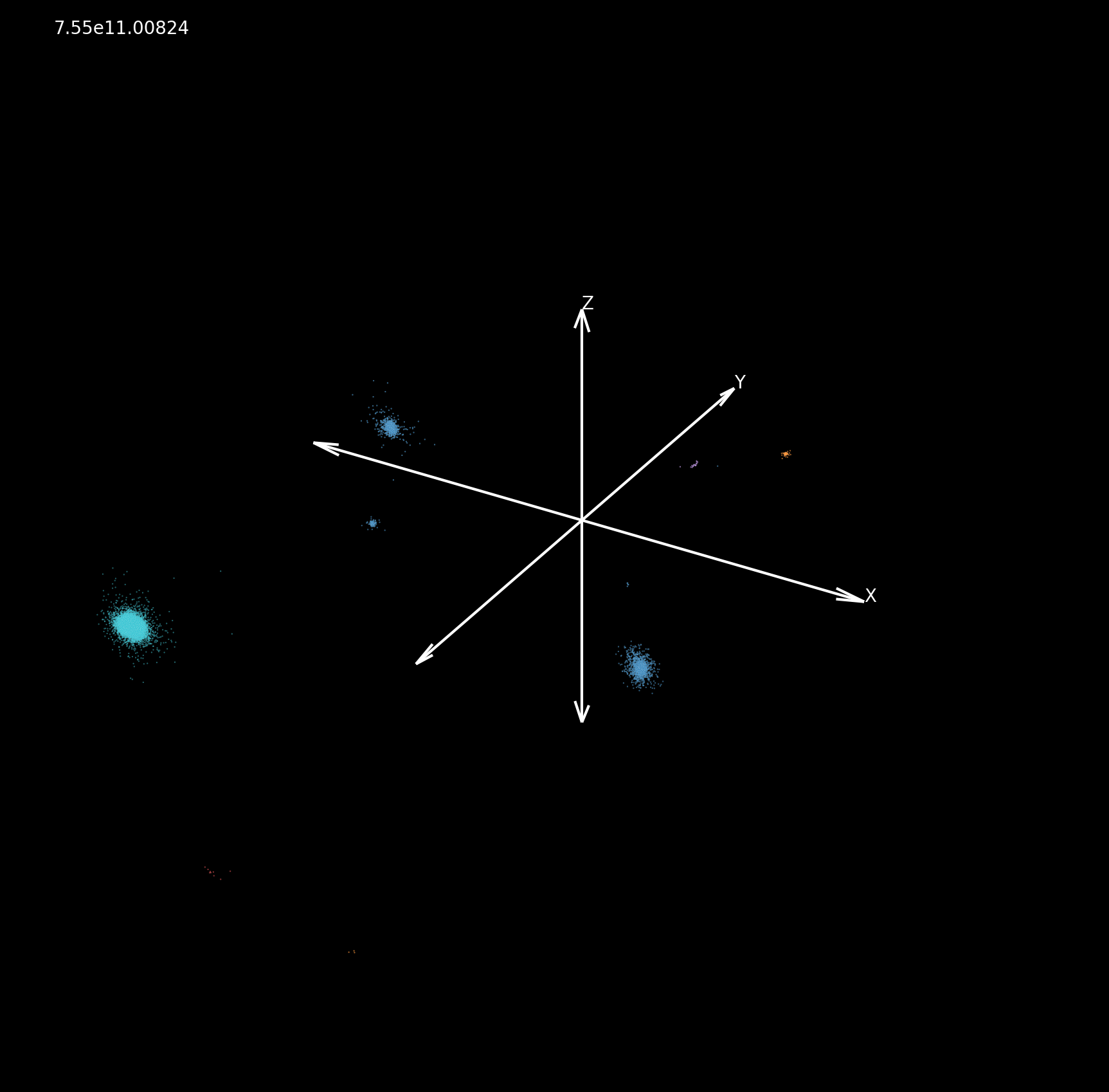}};
    \node at (img6.west) [xshift = -0.075\textwidth, align=center] {7.55e11.00824\\$\sim 2.7$ Gyr};
    %
    %
    \node (img7) at (img1.east) [xshift = 0.245\textwidth]
    {\includegraphics[width=0.5\textwidth,trim={10mm 15mm 11mm 10mm},clip]{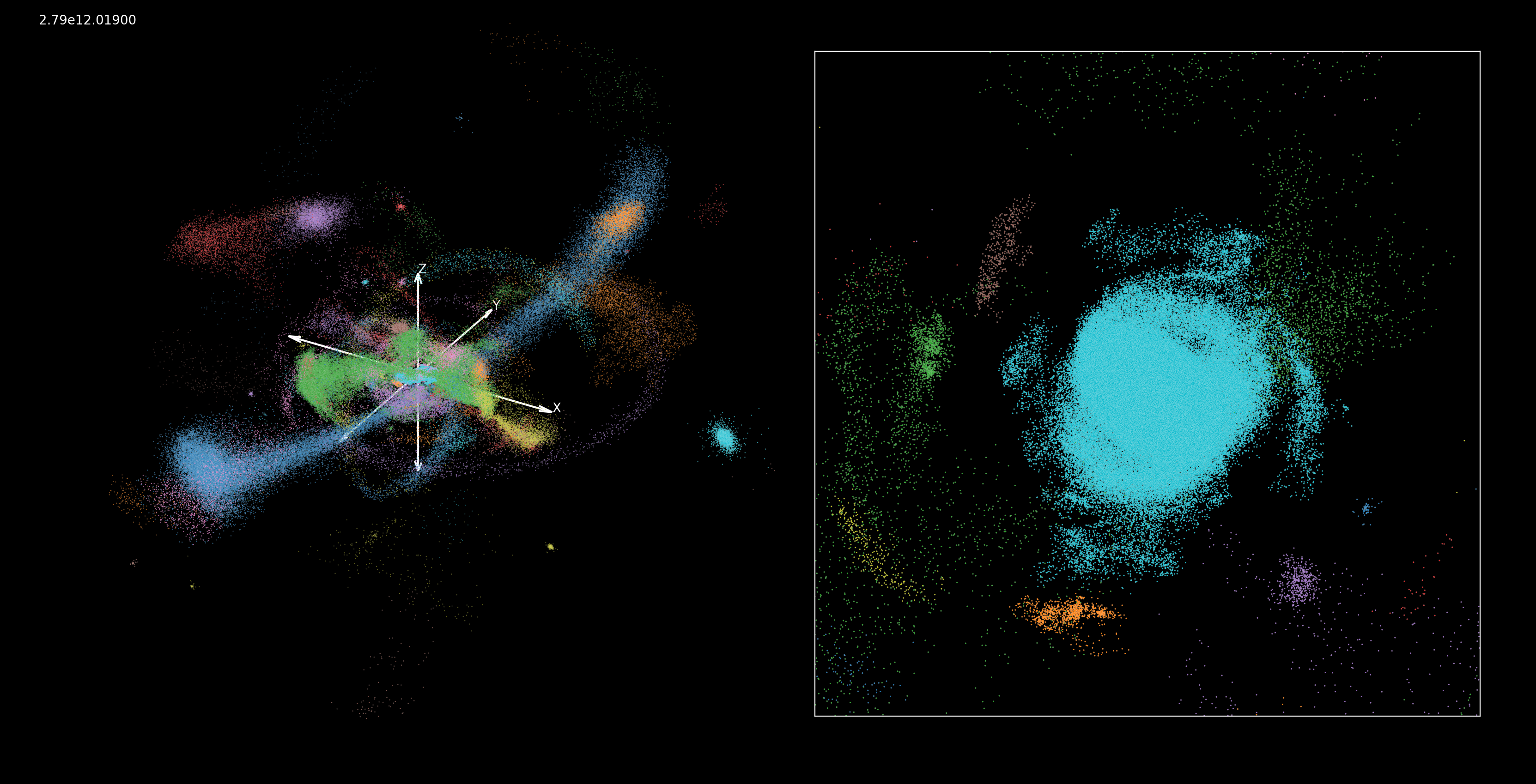}};
    \node at (img7.north) [yshift = 0.01\textwidth] {\large \textbf{FuzzyCat $\circ$ AstroLink}};
    \node (txt1) at (img7.north west) [xshift = 0.18\textwidth, yshift = -0.205\textwidth] {\footnotesize \textcolor{white}{streams}};
    \draw[->, white, thick] (txt1) -- +(-0.095\textwidth, 0.04\textwidth);
    \draw[->, white, thick] (txt1) -- +(-0.042\textwidth, 0.04\textwidth);
    \draw[->, white, thick] (txt1) -- +(-0.005\textwidth, 0.05675\textwidth);
    \draw[->, white, thick] (txt1) -- +(0.02\textwidth, 0.0525\textwidth);
    \node (txt2) at (img7.north west) [xshift = 0.125\textwidth, yshift = -0.03\textwidth] {\footnotesize \textcolor{white}{progenitor}};
    \draw[->, white, thick] (txt2) -- +(0.075\textwidth, -0.04\textwidth);
    \node (img8) at (img2.east) [xshift = 0.245\textwidth]
    {\includegraphics[width=0.5\textwidth,trim={10mm 15mm 11mm 10mm},clip]{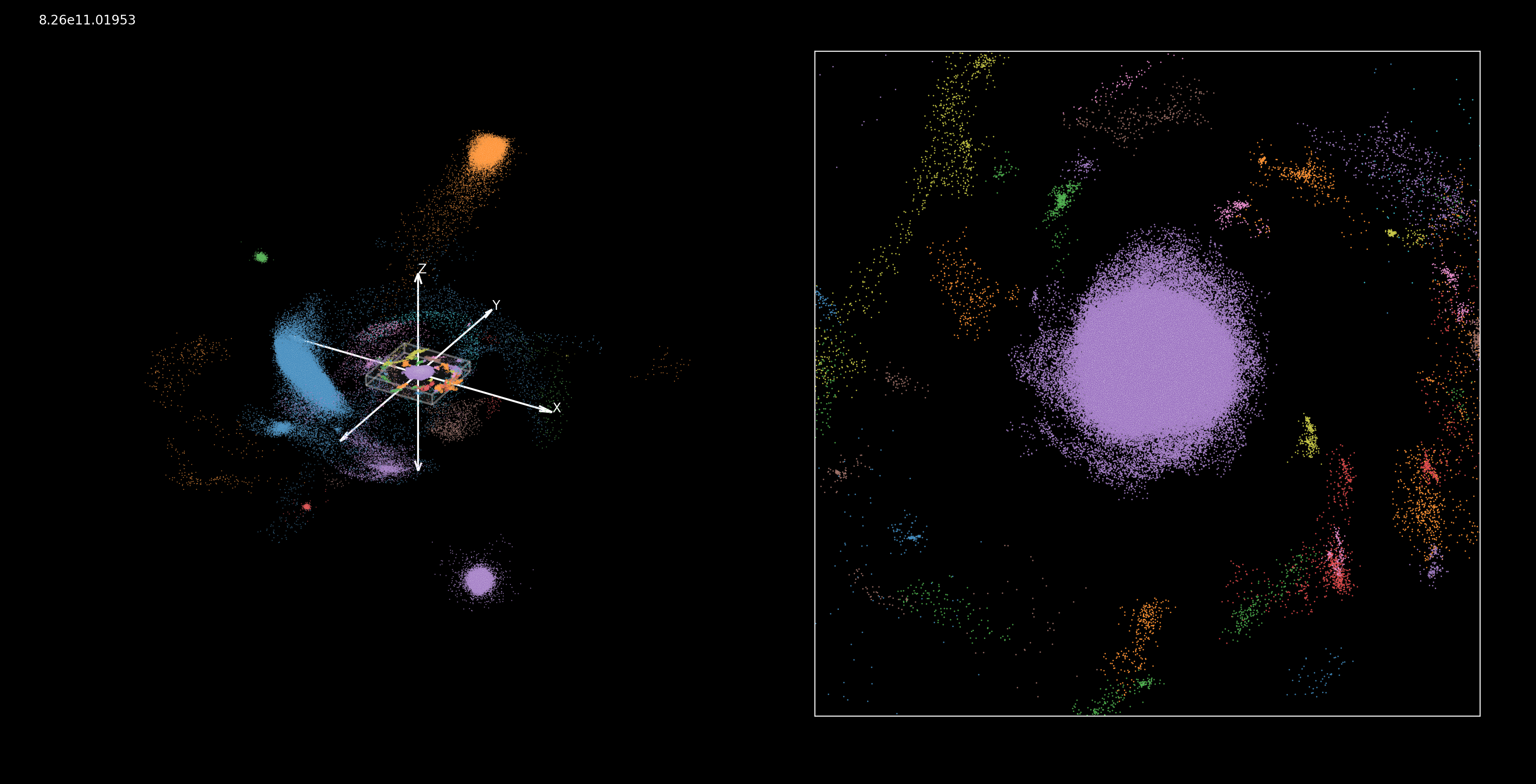}};
    \node (txt3) at (img8.north west) [xshift = 0.23\textwidth, yshift = -0.12\textwidth, align=center, text width = 1.3cm] {\footnotesize \textcolor{white}{dwarf galaxies}};
    \draw[->, white, thick] (txt3) -- +(-0.06\textwidth, 0.0675\textwidth);
    \draw[->, white, thick] (txt3) -- +(-0.065\textwidth, -0.07\textwidth);
    \node (txt4) at (img8.north west) [xshift = 0.34\textwidth, yshift = -0.2\textwidth, align=center, text width = 1.6cm] {\footnotesize \textcolor{white}{star-forming regions}};
    \draw[->, white, thick] (txt4) -- +(-0.01\textwidth, 0.09\textwidth);
    \draw[->, white, thick] (txt4) -- +(0.015\textwidth, 0.12\textwidth);
    \draw[->, white, thick] (txt4) -- +(0.135\textwidth, 0.1\textwidth);
    \draw[->, white, thick] (txt4) -- +(0.09\textwidth, 0.02\textwidth);
    \node (img9) at (img3.east) [xshift = 0.245\textwidth]
    {\includegraphics[width=0.5\textwidth,trim={10mm 15mm 11mm 10mm},clip]{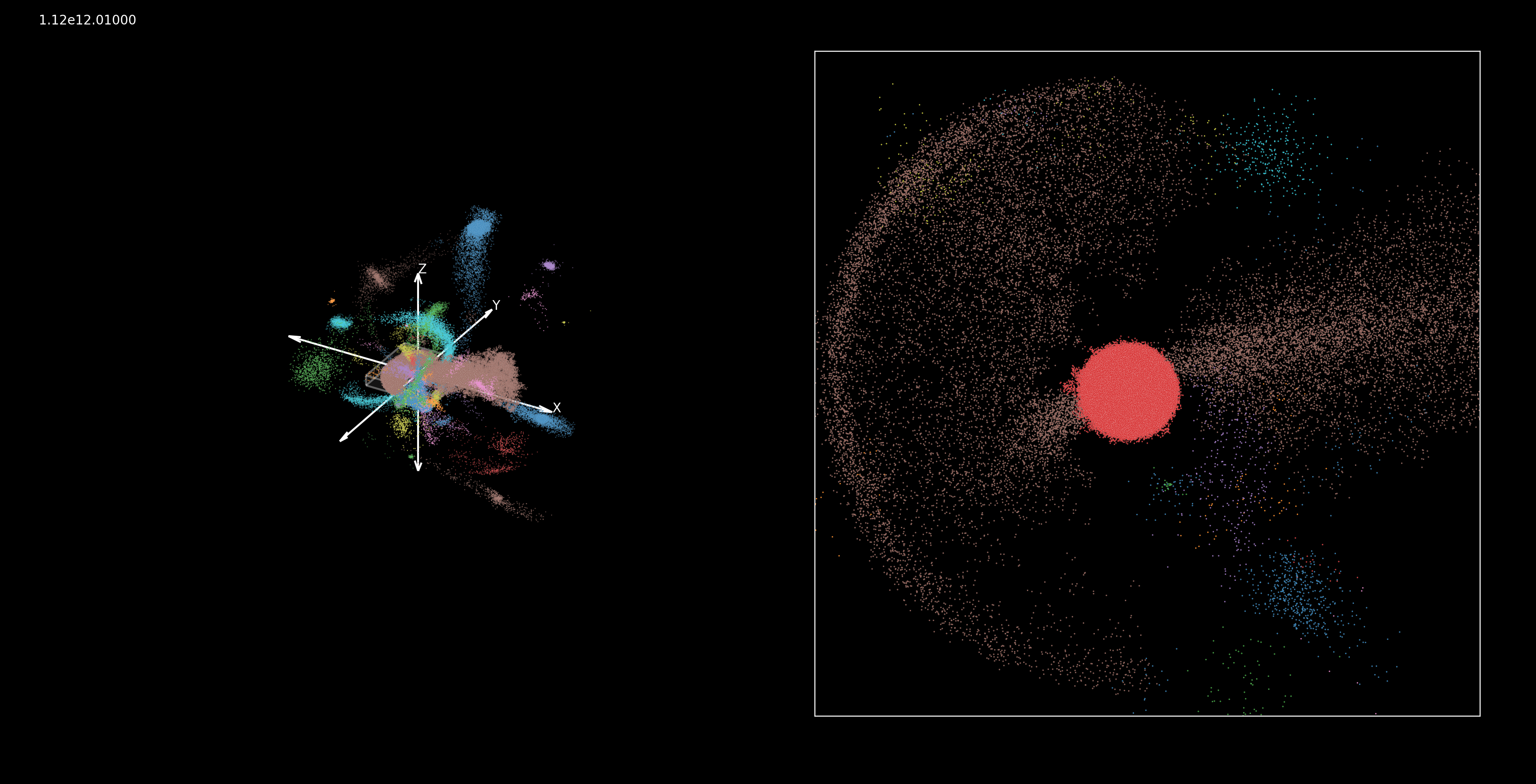}};
    \node (txt5) at (img9.north west) [xshift = 0.245\textwidth, yshift = -0.04\textwidth] {\footnotesize \textcolor{white}{shell}};
    \draw[->, white, thick] (txt5) -- +(0.04\textwidth, -0.04\textwidth);
    \node (txt6) at (img9.north west) [xshift = 0.3825\textwidth, yshift = -0.195\textwidth] {\footnotesize \textcolor{white}{bulge}};
    \draw[->, white, thick] (txt6) -- +(0.0\textwidth, 0.045\textwidth);
    \node (img10) at (img4.east) [xshift = 0.245\textwidth]
    {\includegraphics[width=0.5\textwidth,trim={10mm 15mm 11mm 10mm},clip]{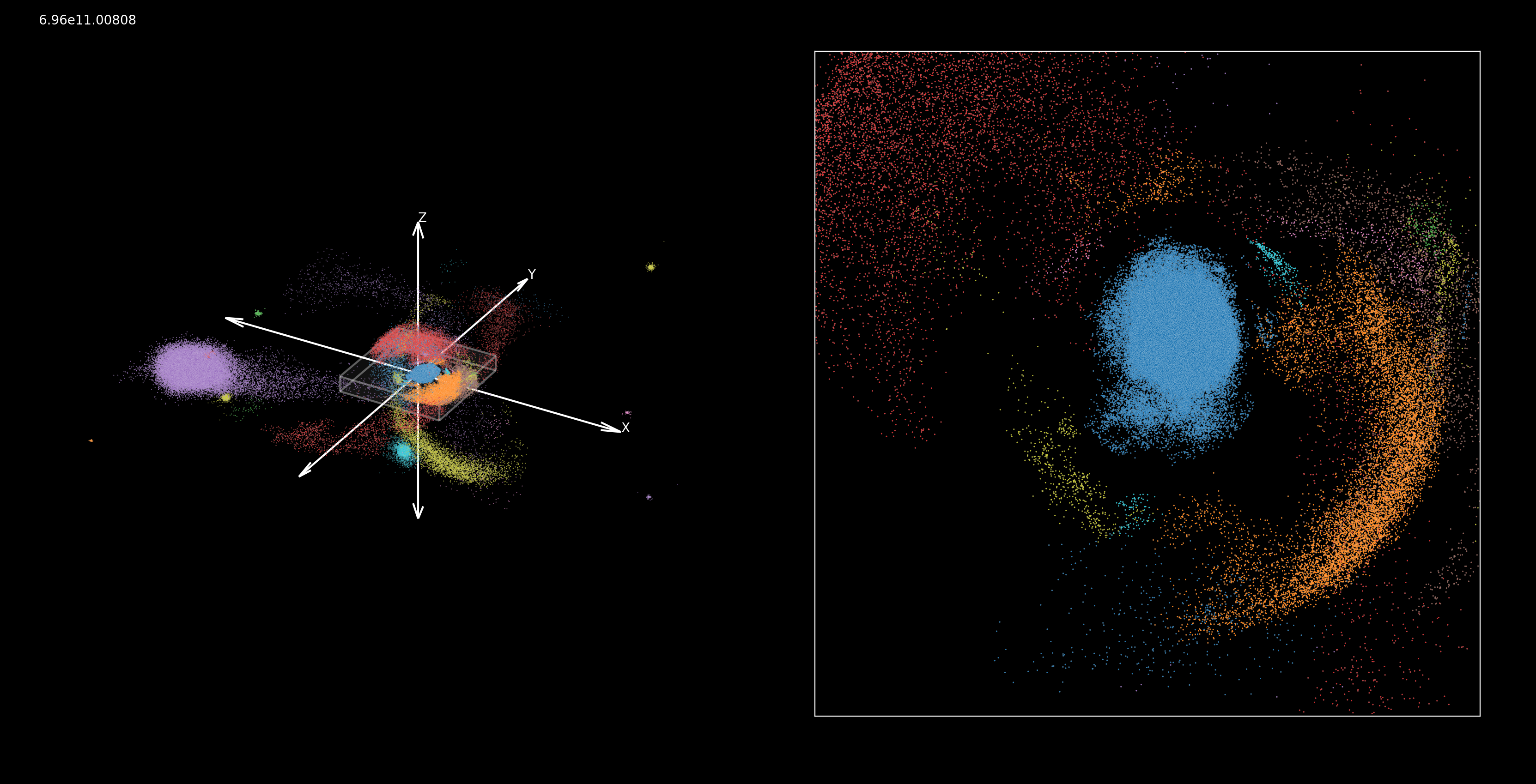}};
    \node (txt7) at (img10.north west) [xshift = 0.075\textwidth, yshift = -0.215\textwidth, align=center, text width = 2cm] {\footnotesize \textcolor{white}{disrupting dwarf galaxy}};
    \draw[->, white, thick] (txt7) -- +(-0.01\textwidth, 0.083\textwidth);
    \node (img11) at (img5.east) [xshift = 0.245\textwidth]
    {\includegraphics[width=0.5\textwidth,trim={10mm 15mm 11mm 10mm},clip]{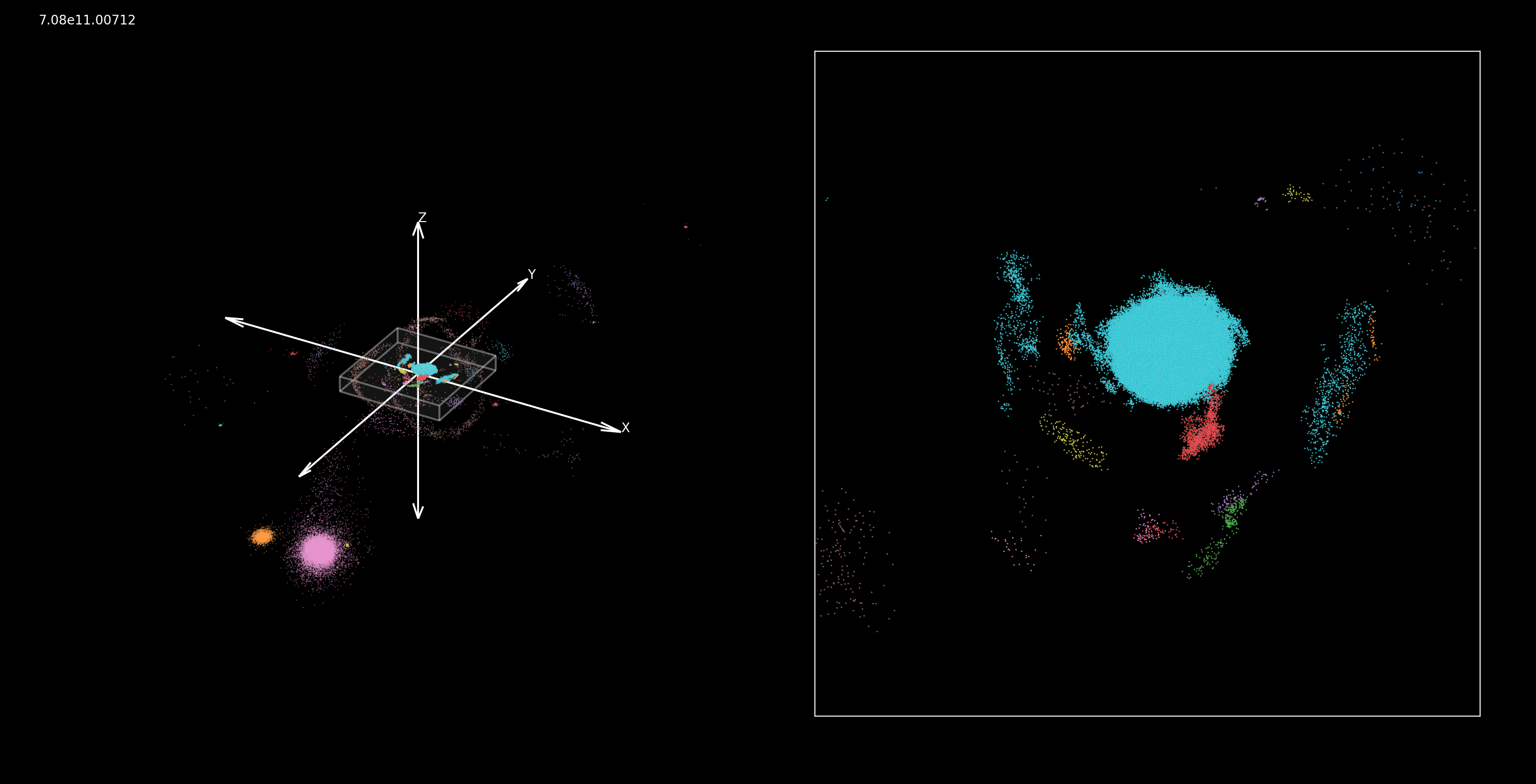}};
    \node (txt8) at (img11.north west) [xshift = 0.195\textwidth, yshift = -0.215\textwidth, align=center, text width = 2.3cm] {\footnotesize \textcolor{white}{thin stream with multiple loops}};
    \draw[->, white, thick] (txt8) -- +(-0.04\textwidth, 0.07\textwidth);
    \node (img12) at (img6.east) [xshift = 0.245\textwidth]
    {\includegraphics[width=0.5\textwidth,trim={10mm 15mm 11mm 10mm},clip]{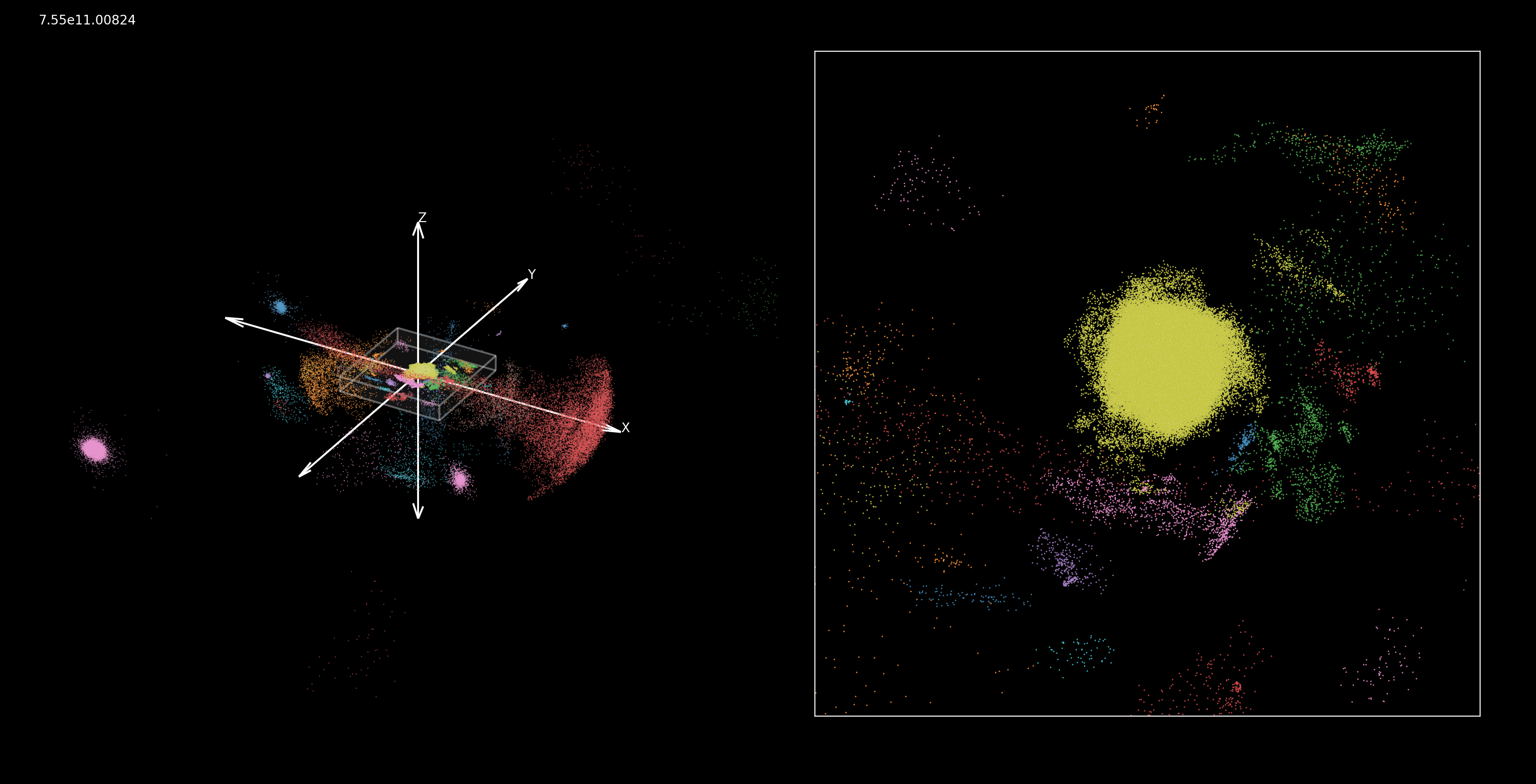}};
    \node (txt9) at (img12.north west) [xshift = 0.21\textwidth, yshift = -0.215\textwidth, align=center, text width = 2.1cm] {\footnotesize \textcolor{white}{multi-shelled tidal debris}};
    \draw[->, white, thick] (txt9) -- +(-0.01\textwidth, 0.06\textwidth);
\end{tikzpicture}
\end{center}
\vspace{-0.4cm}
\caption{Frames from animations (code \& results found \href{https://fuzzycat.readthedocs.io/en/latest/phasetemporalclustering.html\#analysis-let-s-visualise-the-results}{\textcolor{blue}{\textbf{here}}}) of clusterings produced by \texttt{AHF} (left) and the \fuzzycat\ $\circ$ \astrolink\ pipeline (right). Each row shows the snapshot's file name (i.e. \texttt{galaxy\_mass.snapshot\_number}) and look-back time. Each panel depicts a full 3D projection (axes extend $100\ \kpc$ from the galactic centre), and for our approach, a zoomed-in top-down view of the galactic disk (i.e. everything within the central white prism of width $50\ \kpc$ and height $10\ \kpc$).}
\label{fig:phasetemporalclusterings}
\vspace{-0.5cm}
\end{figure}

It is worth clarifying that \fuzzycat\ is never provided any knowledge of the feature space representation of the object-based input data set nor of the underlying change process that acts upon it -- i.e. it makes no assumptions about \textit{why} clusters of these objects exist nor about \textit{how} they may change between any two clusterings. It is easier to see why \fuzzycat\ can be applied to a stochastically changing clustering (as in Sec. \ref{sec:simple_case}) as opposed to a temporally evolving one (as in Sec. \ref{fig:phasetemporalclusterings}) -- however the strong temporal correlation between consecutive snapshots produces a statistically significant clustering signal resulting in physically meaningful cluster tracking. We refer the reader to the \fuzzycat\ ReadTheDocs page \citep{FuzzyCatReadTheDocs2024} for more details.


\begin{wrapfigure}[16]{R}{0.675\textwidth}
    \centering
    \vspace{-1.9cm}
    \includegraphics[width=0.33\textwidth,trim={25.9315mm 22.1405mm 20mm 23mm},clip]{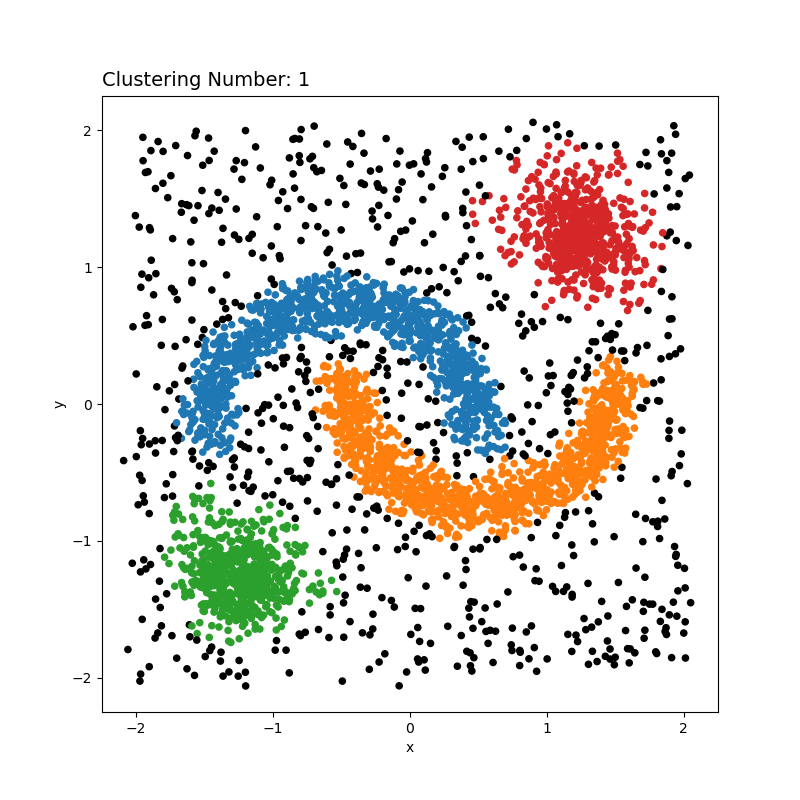}
    \includegraphics[width=0.33\textwidth,trim={25.9315mm 22.1405mm 20mm 23mm},clip]{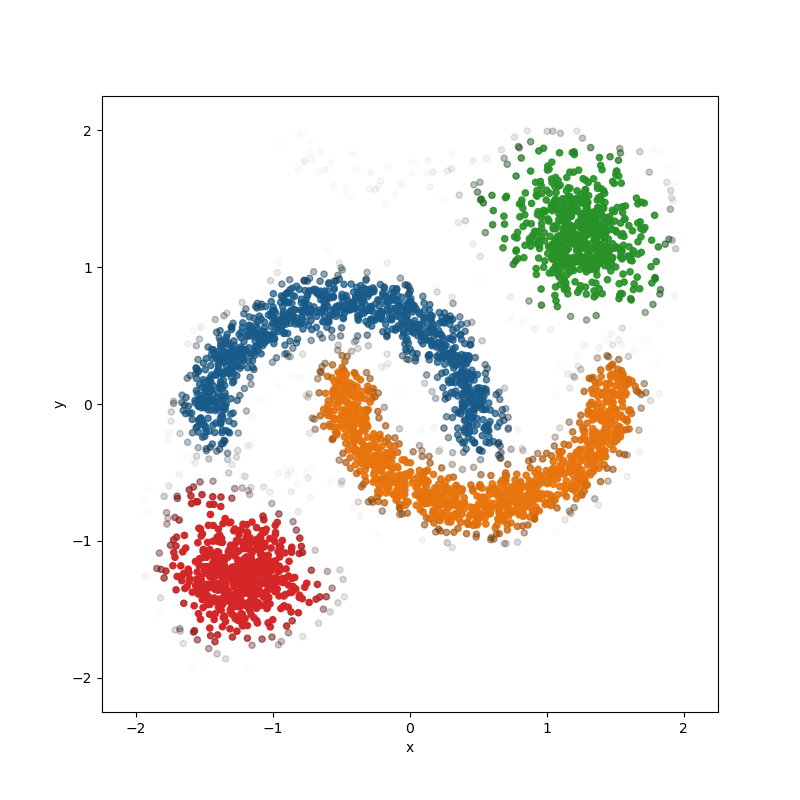}
    \vspace{-0.25cm}
    \caption{The \fuzzycat\ $\circ$ \astrolink\ pipeline applied to uncertain toy data (code \& results found \href{https://fuzzycat.readthedocs.io/en/latest/intro.html\#basic-usage}{\textcolor{blue}{\textbf{here}}}). \textit{Left}: A random sample of the data with coloured points belonging to \astrolink\ clusters. \textit{Right}: The resultant fuzzy clusters found by \fuzzycat\ after $100$ resamplings/clusterings of the data with the opacity and colour of points representing the membership function (in this case a probability) of a data point to belong to a particular fuzzy cluster.}
    \label{fig:simple_case}
    \vspace{-1.8cm}
\end{wrapfigure}

\section{A simple use case} \label{sec:simple_case}
We first take a detour from the focus of this work and demonstrate the versatility of this pipeline by applying it to a 2D toy data set whereby the underlying process that changes the data is stochastic resampling due to the effect of uncertainties. In this exercise, each data point's uncertainty profile is a 2D Gaussian distribution with identical covariance matrices equal to $\sigma^2 I$ where $\sigma = 0.05$. The data has been resampled and clustered $100$ times by \astrolink\ and then \fuzzycat\ is applied to these resampled clusterings (with both algorithms using their default settings). Fig. \ref{fig:simple_case} depicts the results, where we see that the effect of stochastically resampling from the uncertainty profiles of the data points is to give fuzzy boundaries to the \astrolink\ clusters. Although we don't explore this function further in this work, it shows that the \fuzzycat\ $\circ$ \astrolink\ pipeline is capable of propagating uncertainties into clusters. Such a function would also be highly beneficial for studying galaxy formation and evolution in the context of observational data, as well as for other areas of the physical sciences where uncertainties are inherent to the data used.

\section{Applications to NIHAO-UHD galaxies and comparison with AHF} \label{sec:phasetemporalclustering}

We now apply the \fuzzycat\ $\circ$ \astrolink\ pipeline to a set of six simulated galaxies from the NIHAO-UHD suite \citep{Buck2020a} -- detailed in Sec. \ref{sec:nihao_uhd}. We first apply \astrolink\ with its default settings to the 6D position-velocity feature space of the stellar particles in each snapshot of each galaxy. We then provide the resultant clusterings to \fuzzycat\ where we choose the \texttt{minStability} hyperparameter such that the resultant fuzzy clusters exist for $\geq230$ Myr (approximately the period of the Sun's orbit within the Milky Way). So as to draw a comparison to current methods, we also apply \texttt{Amiga's Halo Finder} (\texttt{AHF}) to the same set of galaxies -- see Sec. \ref{sec:ahf} for details on this code.

Fig. \ref{fig:phasetemporalclusterings} depicts various frames from the animations generated via both \texttt{AHF} and the \fuzzycat\ $\circ$ \astrolink\ pipeline. Among the structures extracted by our approach are; dwarf galaxies, infalling groups, stellar streams (and their progenitors), stellar shells, galactic bulges, and star-forming regions. By comparison, traditional approaches are not able to find most of this structure beyond a subset of which is (or is mostly) self-bound -- this can be seen with the corresponding results from \texttt{AHF}. Our pipeline therefore reveals much more of the information content of these galaxies, and as such, lends itself as a powerful tool for analysing galaxy formation and evolution in a modern setting. We refer the reader to the \fuzzycat\ ReadTheDocs page \citep{FuzzyCatReadTheDocs2024} for a coded tutorial of this work as well as animations of the phase-temporal clusterings for each of the six galaxies by each of the codes.



\subsection{The NIHAO-UHD suite of cosmological hydrodynamical simulations} \label{sec:nihao_uhd}

The NIHAO-UHD suite \citep{Buck2020a} is an ultra-high resolution subset of the Numerical Investigation of a Hundred Astronomical Objects (NIHAO) simulation suite \citep{Wang2015} which assumes cosmological parameters from the 2015 \citet{Planck2014}. These galaxies are chosen to reflect the most MW-like galaxies in terms of mass, size and disk properties. Parts of the simulation suite have previously been used to study the build-up of MW's peanut-shaped bulge \citep{Buck2018, Buck2019b}, investigate the stellar bar properties \citep{Hilmi2020}, infer the MW's dark halo spin \citep{Obreja2022}, study the dwarf galaxy inventory of MW mass galaxies \citep{Buck2019a, Buck2023}, and investigate the age-metallicity relation of MW disk stars \citep{Lu2022b} including the chemical bimodality of disk stars \citep{Buck2020b,Buck2021b}, their abundances \citep{Lu2022a} and the origin of very metal-poor stars inside the stellar disk \citep{Sestito2021}. Because of their realistic cosmology, complex hydrodynamical nature, and advanced physical realism, the NIHAO-UHD galaxies serve as an excellent probe for studying galaxy formation and evolution -- as well as to showcase the novel capabilities of our pipeline.

\subsection{Amiga's Halo Finder} \label{sec:ahf}

As a comparison to our approach, we apply the halo finder \texttt{AHF2} \citep{Gill2004, Knollmann2009} to each snapshot of each simulated galaxy. It works by recursively refining a grid in a top-down manner to identify spatial regions within the simulation box that meet a certain overdensity threshold ($200$ times the critical density of the Universe, in this case). Once particles are identified as belonging to such regions, an iterative unbinding procedure is performed to remove all particles whose velocity exceeds the escape velocity at that particle's position within the given halo (this assumes a spherically symmetric density profile for the halo). The recursive and iterative nature of this algorithm yields a hierarchy of (sub)haloes -- bounded groups with an overdensity above a threshold. With the haloes of each snapshot now found, temporal merger trees are calculated using the \texttt{AHF} analysis tool, \texttt{MergerTree}, which traces the particle IDs of all particles throughout the snapshots and identifies all progenitors of a given halo. The animations, from which the frames in Fig. \ref{fig:phasetemporalclusterings} are taken, plot the star particles from each \texttt{AHF} halo with a colour that is passed down from its `father' halo (its most similar progenitor from the previous snapshot). In this sense, these animations are the \texttt{AHF} analogue of those produced using the results of the \fuzzycat\ $\circ$ \astrolink\ pipeline.

\section{Conclusion and outlook}

In this work, we have demonstrated the effectiveness of the \fuzzycat\ $\circ$ \astrolink\ pipeline as a novel unsupervised machine learning approach -- particularly as a tool for analysing simulated galaxies in the context of galaxy formation and evolution. By applying our pipeline to the NIHAO-UHD suite, we have shown that it can successfully identify a diverse range of astrophysical structures that traditional halo finder (+ merger tree) methods do not -- capturing transient and tidally disrupted structures that are often overlooked in conventional analyses. As such, it provides the means to a more comprehensive understanding of galaxy formation and evolution.

The ability of the \fuzzycat\ $\circ$ \astrolink\ pipeline to adapt to data with underlying processes, such as stochastic variations and temporal evolution, positions it as a powerful tool for future studies in astrophysics and in other fields where data is fuzzy, dynamic, and complex. By overcoming the limitations of existing methods, our approach offers its user a more flexible and detailed examination of the hierarchical and multifaceted nature of astrophysical structures. We now intend to improve upon our pipeline through massive parallel (re-)implementations of \astrolink\ and \fuzzycat\ -- opening application avenues to a broader range of cosmological simulations and observational data sets, thereby enhancing our understanding of our Universe's structure at multiple scales.

\section*{Broader impact statement}
The authors are not aware of any immediate ethical or societal implications of this work. This work purely aims to aid scientific research and proposes a method of using a pipeline of clustering algorithms to learn about galaxy formation and evolution.

\begin{ack}
This work is funded by the Carl-Zeiss-Stiftung through the NEXUS program.
\end{ack}

\bibliography{references}

\begin{thebibliography}{45}
\providecommand{\natexlab}[1]{#1}
\providecommand{\url}[1]{\texttt{#1}}
\expandafter\ifx\csname urlstyle\endcsname\relax
  \providecommand{\doi}[1]{doi: #1}\else
  \providecommand{\doi}{doi: \begingroup \urlstyle{rm}\Url}\fi

\bibitem[Ankerst et~al.(1999)Ankerst, Breunig, Kriegel, and Sander]{Ankerst1999}
Mihael Ankerst, Markus~M Breunig, Hans-Peter Kriegel, and Jörg Sander.
\newblock Optics: ordering points to identify the clustering structure.
\newblock In \emph{ACM Sigmod record}, volume~28, pages 49--60. ACM, 1999.
\newblock ISBN 1581130848.
\newblock \doi{10.1145/304182.304187}.

\bibitem[Avila et~al.(2014)Avila, Knebe, Pearce, Schneider, Srisawat, Thomas, Behroozi, Elahi, Han, Mao, Onions, Rodriguez-Gomez, and Tweed]{Avila2014}
Santiago Avila, Alexander Knebe, Frazer~R. Pearce, Aurel Schneider, Chaichalit Srisawat, Peter~A. Thomas, Peter Behroozi, Pascal~J. Elahi, Jiaxin Han, Yao-Yuan Mao, Julian Onions, Vicente Rodriguez-Gomez, and Dylan Tweed.
\newblock {SUSSING MERGER TREES: the influence of the halo finder}.
\newblock \emph{Monthly Notices of the Royal Astronomical Society}, 441\penalty0 (4):\penalty0 3488--3501, 05 2014.
\newblock ISSN 0035-8711.
\newblock \doi{10.1093/mnras/stu799}.
\newblock URL \url{https://doi.org/10.1093/mnras/stu799}.

\bibitem[Behroozi et~al.(2015)Behroozi, Knebe, Pearce, Elahi, Han, Lux, Mao, Muldrew, Potter, and Srisawat]{Behroozi2015}
Peter Behroozi, Alexander Knebe, Frazer~R. Pearce, Pascal Elahi, Jiaxin Han, Hanni Lux, Yao-Yuan Mao, Stuart~I. Muldrew, Doug Potter, and Chaichalit Srisawat.
\newblock {Major mergers going Notts: challenges for modern halo finders}.
\newblock \emph{Monthly Notices of the Royal Astronomical Society}, 454\penalty0 (3):\penalty0 3020--3029, 10 2015.
\newblock ISSN 0035-8711.
\newblock \doi{10.1093/mnras/stv2046}.
\newblock URL \url{https://doi.org/10.1093/mnras/stv2046}.

\bibitem[Behroozi et~al.(2012)Behroozi, Wechsler, and Wu]{Behroozi2012}
Peter~S. Behroozi, Risa~H. Wechsler, and Hao-Yi Wu.
\newblock The rockstar phase-space temporal halo finder and the velocity offsets of cluster cores.
\newblock \emph{The Astrophysical Journal}, 762\penalty0 (2):\penalty0 109, 2012.
\newblock ISSN 0004-637X 1538-4357.
\newblock \doi{10.1088/0004-637x/762/2/109}.
\newblock URL \url{http://dx.doi.org/10.1088/0004-637X/762/2/109}.

\bibitem[{Buck} et~al.(2019{\natexlab{a}}){Buck}, {Ness}, {Obreja}, {Macci{\`o}}, and {Dutton}]{Buck2019b}
T.~{Buck}, M.~{Ness}, A.~{Obreja}, A.~V. {Macci{\`o}}, and A.~A. {Dutton}.
\newblock {Stars behind Bars II: A Cosmological Formation Scenario for the Milky Way's Central Stellar Structure}.
\newblock \emph{The Astrophysical Journal}, 874:\penalty0 67, March 2019{\natexlab{a}}.
\newblock \doi{10.3847/1538-4357/aaffd0}.

\bibitem[{Buck}(2020)]{Buck2020b}
Tobias {Buck}.
\newblock {On the origin of the chemical bimodality of disc stars: a tale of merger and migration}.
\newblock \emph{Monthly Notices of the Royal Astronomical Society}, 491\penalty0 (4):\penalty0 5435--5446, February 2020.
\newblock \doi{10.1093/mnras/stz3289}.

\bibitem[{Buck} et~al.(2018){Buck}, {Ness}, {Macci{\`o}}, {Obreja}, and {Dutton}]{Buck2018}
Tobias {Buck}, Melissa~K. {Ness}, Andrea~V. {Macci{\`o}}, Aura {Obreja}, and Aaron~A. {Dutton}.
\newblock {Stars Behind Bars. I. The Milky Way's Central Stellar Populations}.
\newblock \emph{The Astrophyscial Journal}, 861\penalty0 (2):\penalty0 88, July 2018.
\newblock \doi{10.3847/1538-4357/aac890}.

\bibitem[{Buck} et~al.(2019{\natexlab{b}}){Buck}, {Macci{\`o}}, {Dutton}, {Obreja}, and {Frings}]{Buck2019a}
Tobias {Buck}, Andrea~V. {Macci{\`o}}, Aaron~A. {Dutton}, Aura {Obreja}, and Jonas {Frings}.
\newblock {NIHAO XV: the environmental impact of the host galaxy on galactic satellite and field dwarf galaxies}.
\newblock \emph{Monthly Notices of the Royal Astronomical Society}, 483\penalty0 (1):\penalty0 1314--1341, February 2019{\natexlab{b}}.
\newblock \doi{10.1093/mnras/sty2913}.

\bibitem[{Buck} et~al.(2020){Buck}, {Obreja}, {Macci{\`o}}, {Minchev}, {Dutton}, and {Ostriker}]{Buck2020a}
Tobias {Buck}, Aura {Obreja}, Andrea~V. {Macci{\`o}}, Ivan {Minchev}, Aaron~A. {Dutton}, and Jeremiah~P. {Ostriker}.
\newblock {NIHAO-UHD: the properties of MW-like stellar discs in high-resolution cosmological simulations}.
\newblock \emph{Monthly Notices of the Royal Astronomical Society}, 491\penalty0 (3):\penalty0 3461--3478, January 2020.
\newblock \doi{10.1093/mnras/stz3241}.

\bibitem[{Buck} et~al.(2021){Buck}, {Rybizki}, {Buder}, {Obreja}, {Macci{\`o}}, {Pfrommer}, {Steinmetz}, and {Ness}]{Buck2021b}
Tobias {Buck}, Jan {Rybizki}, Sven {Buder}, Aura {Obreja}, Andrea~V. {Macci{\`o}}, Christoph {Pfrommer}, Matthias {Steinmetz}, and Melissa {Ness}.
\newblock {The challenge of simultaneously matching the observed diversity of chemical abundance patterns in cosmological hydrodynamical simulations}.
\newblock \emph{Monthly Notices of the Royal Astronomical Society}, 508\penalty0 (3):\penalty0 3365--3387, December 2021.
\newblock \doi{10.1093/mnras/stab2736}.

\bibitem[{Buck} et~al.(2023){Buck}, {Obreja}, {Ratcliffe}, {Lu}, {Minchev}, and {Macci{\`o}}]{Buck2023}
Tobias {Buck}, Aura {Obreja}, Bridget {Ratcliffe}, Yuxi(Lucy) {Lu}, Ivan {Minchev}, and Andrea~V. {Macci{\`o}}.
\newblock {The impact of early massive mergers on the chemical evolution of Milky Way-like galaxies: insights from NIHAO-UHD simulations}.
\newblock \emph{Monthly Notices of the Royal Astronomical Society}, 523\penalty0 (1):\penalty0 1565--1576, July 2023.
\newblock \doi{10.1093/mnras/stad1503}.

\bibitem[Campello et~al.(2015)Campello, Moulavi, Zimek, and Sander]{Campello2015}
Ricardo J. G.~B. Campello, Davoud Moulavi, Arthur Zimek, and J\"{o}rg Sander.
\newblock Hierarchical density estimates for data clustering, visualization, and outlier detection.
\newblock \emph{ACM Trans. Knowl. Discov. Data}, 10\penalty0 (1), July 2015.
\newblock ISSN 1556-4681.
\newblock \doi{10.1145/2733381}.
\newblock URL \url{https://doi.org/10.1145/2733381}.

\bibitem[Davis et~al.(1985)Davis, Efstathiou, Frenk, and White]{Davis1985}
Marc Davis, George Efstathiou, Carlos~S Frenk, and Simon~DM White.
\newblock The evolution of large-scale structure in a universe dominated by cold dark matter.
\newblock \emph{The Astrophysical Journal}, 292:\penalty0 371--394, 1985.
\newblock \doi{10.1086/163168}.

\bibitem[Elahi et~al.(2013)Elahi, Han, Lux, Ascasibar, Behroozi, Knebe, Muldrew, Onions, and Pearce]{Elahi2013}
Pascal~J. Elahi, Jiaxin Han, Hanni Lux, Yago Ascasibar, Peter Behroozi, Alexander Knebe, Stuart~I. Muldrew, Julian Onions, and Frazer Pearce.
\newblock {Streams going Notts: the tidal debris finder comparison project}.
\newblock \emph{Monthly Notices of the Royal Astronomical Society}, 433\penalty0 (2):\penalty0 1537--1555, 06 2013.
\newblock ISSN 0035-8711.
\newblock \doi{10.1093/mnras/stt825}.
\newblock URL \url{https://doi.org/10.1093/mnras/stt825}.

\bibitem[Epanechnikov(1969)]{Epanechnikov1969}
Vassiliy~A Epanechnikov.
\newblock Non-parametric estimation of a multivariate probability density.
\newblock \emph{Theory of Probability \& Its Applications}, 14\penalty0 (1):\penalty0 153--158, 1969.

\bibitem[Ester et~al.(1996)Ester, Kriegel, Sander, and Xu]{Ester1996}
Martin Ester, Hans-Peter Kriegel, Jörg Sander, and Xiaowei Xu.
\newblock A density-based algorithm for discovering clusters in large spatial databases with noise.
\newblock In \emph{Kdd}, volume~96, pages 226--231, 1996.

\bibitem[{Gill} et~al.(2004){Gill}, {Knebe}, and {Gibson}]{Gill2004}
Stuart P.~D. {Gill}, Alexander {Knebe}, and Brad~K. {Gibson}.
\newblock {The evolution of substructure - I. A new identification method}.
\newblock \emph{Monthly Notices of the Royal Astronomical Society}, 351\penalty0 (2):\penalty0 399--409, June 2004.
\newblock \doi{10.1111/j.1365-2966.2004.07786.x}.

\bibitem[Han et~al.(2017)Han, Cole, Frenk, Benitez-Llambay, and Helly]{Han2018}
Jiaxin Han, Shaun Cole, Carlos~S. Frenk, Alejandro Benitez-Llambay, and John Helly.
\newblock {hbt+: an improved code for finding subhaloes and building merger trees in cosmological simulations}.
\newblock \emph{Monthly Notices of the Royal Astronomical Society}, 474\penalty0 (1):\penalty0 604--617, 10 2017.
\newblock ISSN 0035-8711.
\newblock \doi{10.1093/mnras/stx2792}.
\newblock URL \url{https://doi.org/10.1093/mnras/stx2792}.

\bibitem[{Hilmi} et~al.(2020){Hilmi}, {Minchev}, {Buck}, {Martig}, {Quillen}, {Monari}, {Famaey}, {de Jong}, {Laporte}, {Read}, {Sanders}, {Steinmetz}, and {Wegg}]{Hilmi2020}
T.~{Hilmi}, I.~{Minchev}, T.~{Buck}, M.~{Martig}, A.~C. {Quillen}, G.~{Monari}, B.~{Famaey}, R.~S. {de Jong}, C.~F.~P. {Laporte}, J.~{Read}, J.~L. {Sanders}, M.~{Steinmetz}, and C.~{Wegg}.
\newblock {Fluctuations in galactic bar parameters due to bar-spiral interaction}.
\newblock \emph{Monthly Notices of the Royal Astronomical Society}, 497\penalty0 (1):\penalty0 933--955, September 2020.
\newblock \doi{10.1093/mnras/staa1934}.

\bibitem[Knebe et~al.(2011)Knebe, Knollmann, Muldrew, Pearce, Aragon-Calvo, Ascasibar, Behroozi, Ceverino, Colombi, Diemand, Dolag, Falck, Fasel, Gardner, Gottlöber, Hsu, Iannuzzi, Klypin, Lukić, Maciejewski, McBride, Neyrinck, Planelles, Potter, Quilis, Rasera, Read, Ricker, Roy, Springel, Stadel, Stinson, Sutter, Turchaninov, Tweed, Yepes, and Zemp]{Knebe2011}
Alexander Knebe, Steffen~R. Knollmann, Stuart~I. Muldrew, Frazer~R. Pearce, Miguel~Angel Aragon-Calvo, Yago Ascasibar, Peter~S. Behroozi, Daniel Ceverino, Stephane Colombi, Juerg Diemand, Klaus Dolag, Bridget~L. Falck, Patricia Fasel, Jeff Gardner, Stefan Gottlöber, Chung-Hsing Hsu, Francesca Iannuzzi, Anatoly Klypin, Zarija Lukić, Michal Maciejewski, Cameron McBride, Mark~C. Neyrinck, Susana Planelles, Doug Potter, Vicent Quilis, Yann Rasera, Justin~I. Read, Paul~M. Ricker, Fabrice Roy, Volker Springel, Joachim Stadel, Greg Stinson, P.~M. Sutter, Victor Turchaninov, Dylan Tweed, Gustavo Yepes, and Marcel Zemp.
\newblock Haloes gone mad14: The halo-finder comparison project.
\newblock \emph{Monthly Notices of the Royal Astronomical Society}, 415\penalty0 (3):\penalty0 2293--2318, 2011.
\newblock ISSN 0035-8711.
\newblock \doi{10.1111/j.1365-2966.2011.18858.x}.
\newblock URL \url{https://doi.org/10.1111/j.1365-2966.2011.18858.x}.

\bibitem[Knebe et~al.(2013)Knebe, Libeskind, Pearce, Behroozi, Casado, Dolag, Dominguez-Tenreiro, Elahi, Lux, Muldrew, et~al.]{Knebe2013a}
Alexander Knebe, Noam~I Libeskind, Frazer Pearce, Peter Behroozi, Javier Casado, Klaus Dolag, Rosa Dominguez-Tenreiro, Pascal Elahi, Hanni Lux, Stuart~I Muldrew, et~al.
\newblock Galaxies going mad: the galaxy-finder comparison project.
\newblock \emph{Monthly Notices of the Royal Astronomical Society}, 428\penalty0 (3):\penalty0 2039--2052, 2013.

\bibitem[{Knollmann} and {Knebe}(2009)]{Knollmann2009}
Steffen~R. {Knollmann} and Alexander {Knebe}.
\newblock {AHF: Amiga's Halo Finder}.
\newblock \emph{The Astrophysical Journal Supplement Series}, 182\penalty0 (2):\penalty0 608--624, June 2009.
\newblock \doi{10.1088/0067-0049/182/2/608}.

\bibitem[Lee et~al.(2014)Lee, Yi, Elahi, Thomas, Pearce, Behroozi, Han, Helly, Jung, Knebe, Mao, Onions, Rodriguez-Gomez, Schneider, Srisawat, and Tweed]{Lee2014}
Jaehyun Lee, Sukyoung~K. Yi, Pascal~J. Elahi, Peter~A. Thomas, Frazer~R. Pearce, Peter Behroozi, Jiaxin Han, John Helly, Intae Jung, Alexander Knebe, Yao-Yuan Mao, Julian Onions, Vicente Rodriguez-Gomez, Aurel Schneider, Chaichalit Srisawat, and Dylan Tweed.
\newblock {Sussing merger trees: the impact of halo merger trees on galaxy properties in a semi-analytic model}.
\newblock \emph{Monthly Notices of the Royal Astronomical Society}, 445\penalty0 (4):\penalty0 4197--4210, 11 2014.
\newblock ISSN 0035-8711.
\newblock \doi{10.1093/mnras/stu2039}.
\newblock URL \url{https://doi.org/10.1093/mnras/stu2039}.

\bibitem[{Lu} et~al.(2022{\natexlab{a}}){Lu}, {Ness}, {Buck}, and {Carr}]{Lu2022b}
Yuxi~Lucy {Lu}, Melissa~K. {Ness}, Tobias {Buck}, and Christopher {Carr}.
\newblock {Turning points in the age-metallicity relations - created by late satellite infall and enhanced by radial migration}.
\newblock \emph{Monthly Notices of the Royal Astronomical Society}, 512\penalty0 (4):\penalty0 4697--4714, June 2022{\natexlab{a}}.
\newblock \doi{10.1093/mnras/stac780}.

\bibitem[{Lu} et~al.(2022{\natexlab{b}}){Lu}, {Ness}, {Buck}, {Zinn}, and {Johnston}]{Lu2022a}
Yuxi~(Lucy) {Lu}, Melissa~K. {Ness}, Tobias {Buck}, Joel~C. {Zinn}, and Kathryn~V. {Johnston}.
\newblock {Similarities behind the high- and low-{\ensuremath{\alpha}} disc: small intrinsic abundance scatter and migrating stars}.
\newblock \emph{Monthly Notices of the Royal Astronomical Society}, 512\penalty0 (2):\penalty0 2890--2910, May 2022{\natexlab{b}}.
\newblock \doi{10.1093/mnras/stac610}.

\bibitem[{Malhan} et~al.(2022){Malhan}, {Ibata}, {Sharma}, {Famaey}, {Bellazzini}, {Carlberg}, {D'Souza}, {Yuan}, {Martin}, and {Thomas}]{Malhan2022}
Khyati {Malhan}, Rodrigo~A. {Ibata}, Sanjib {Sharma}, Benoit {Famaey}, Michele {Bellazzini}, Raymond~G. {Carlberg}, Richard {D'Souza}, Zhen {Yuan}, Nicolas~F. {Martin}, and Guillaume~F. {Thomas}.
\newblock {The Global Dynamical Atlas of the Milky Way Mergers: Constraints from Gaia EDR3-based Orbits of Globular Clusters, Stellar Streams, and Satellite Galaxies}.
\newblock \emph{The Astrophysical Journal}, 926\penalty0 (2):\penalty0 107, February 2022.
\newblock \doi{10.3847/1538-4357/ac4d2a}.

\bibitem[McConnachie et~al.(2018)McConnachie, Ibata, Martin, Ferguson, Collins, Gwyn, Irwin, Lewis, Mackey, Davidge, Arias, Conn, Côté, Crnojevic, Huxor, Penarrubia, Spengler, Tanvir, Valls-Gabaud, Babul, Barmby, Bate, Bernard, Chapman, Dotter, Harris, McMonigal, Navarro, Puzia, Rich, Thomas, and Widrow]{McConnachie2018}
Alan~W. McConnachie, Rodrigo Ibata, Nicolas Martin, Annette M.~N. Ferguson, Michelle Collins, Stephen Gwyn, Mike Irwin, Geraint~F. Lewis, A.~Dougal Mackey, Tim Davidge, Veronica Arias, Anthony Conn, Patrick Côté, Denija Crnojevic, Avon Huxor, Jorge Penarrubia, Chelsea Spengler, Nial Tanvir, David Valls-Gabaud, Arif Babul, Pauline Barmby, Nicholas~F. Bate, Edouard Bernard, Scott Chapman, Aaron Dotter, William Harris, Brendan McMonigal, Julio Navarro, Thomas~H. Puzia, R.~Michael Rich, Guillaume Thomas, and Lawrence~M. Widrow.
\newblock The large-scale structure of the halo of the andromeda galaxy. ii. hierarchical structure in the pan-andromeda archaeological survey.
\newblock \emph{The Astrophysical Journal}, 868\penalty0 (1):\penalty0 55, 2018.
\newblock ISSN 1538-4357.
\newblock \doi{10.3847/1538-4357/aae8e7}.
\newblock URL \url{http://dx.doi.org/10.3847/1538-4357/aae8e7 https://iopscience.iop.org/article/10.3847/1538-4357/aae8e7/pdf}.

\bibitem[McInnes et~al.(2017)McInnes, Healy, and Astels]{McInnes2017}
Leland McInnes, John Healy, and Steve Astels.
\newblock hdbscan: Hierarchical density based clustering.
\newblock \emph{Journal of Open Source Software}, 2\penalty0 (11):\penalty0 205, 2017.
\newblock \doi{10.21105/joss.00205}.
\newblock URL \url{https://doi.org/10.21105/joss.00205}.

\bibitem[{Miro-Carretero} et~al.(2024){Miro-Carretero}, {Gomez-Flechoso}, {Martinez-Delgado}, {Cooper}, {Roca-Fabrega}, {Akhlaghi}, {Pillepich}, {Kuijken}, {Erkal}, {Buck}, {Hellwing}, and {Bose}]{MiroCarretero2024}
Juan {Miro-Carretero}, Maria~A. {Gomez-Flechoso}, David {Martinez-Delgado}, Andrew~P. {Cooper}, Santi {Roca-Fabrega}, Mohammad {Akhlaghi}, Annalisa {Pillepich}, Konrad {Kuijken}, Denis {Erkal}, Tobias {Buck}, Wojciech~A. {Hellwing}, and Sownak {Bose}.
\newblock {Extragalactic Stellar Tidal Streams: Observations meet Simulation}.
\newblock \emph{arXiv e-prints}, art. arXiv:2409.03585, September 2024.
\newblock \doi{10.48550/arXiv.2409.03585}.

\bibitem[{Obreja} et~al.(2022){Obreja}, {Buck}, and {Macci{\`o}}]{Obreja2022}
Aura {Obreja}, Tobias {Buck}, and Andrea~V. {Macci{\`o}}.
\newblock {A first estimate of the Milky Way dark matter halo spin}.
\newblock \emph{Astronomy \& Astrophysics}, 657:\penalty0 A15, January 2022.
\newblock \doi{10.1051/0004-6361/202140983}.

\bibitem[Oliver(2024{\natexlab{a}})]{AstroLinkGithub2024}
William~H Oliver.
\newblock Astrolink, 09 2024{\natexlab{a}}.
\newblock URL \url{https://github.com/william-h-oliver/astrolink}.

\bibitem[Oliver(2024{\natexlab{b}})]{AstroLinkReadTheDocs2024}
William~H Oliver.
\newblock Astrolink, 09 2024{\natexlab{b}}.
\newblock URL \url{https://astrolink.readthedocs.io/en/latest/}.

\bibitem[Oliver(2024{\natexlab{c}})]{FuzzyCatGithub2024}
William~H Oliver.
\newblock Fuzzycat, 09 2024{\natexlab{c}}.
\newblock URL \url{https://github.com/william-h-oliver/fuzzycat}.

\bibitem[Oliver(2024{\natexlab{d}})]{FuzzyCatReadTheDocs2024}
William~H Oliver.
\newblock Fuzzycat, 09 2024{\natexlab{d}}.
\newblock URL \url{https://fuzzycat.readthedocs.io/en/latest/}.

\bibitem[Oliver et~al.(2021)Oliver, Elahi, Lewis, and Power]{Oliver2020}
William~H Oliver, Pascal~J Elahi, Geraint~F Lewis, and Chris Power.
\newblock {The Hierarchical Structure of Galactic Haloes: Classification and Characterization with Halo-OPTICS}.
\newblock \emph{Monthly Notices of the Royal Astronomical Society}, 501\penalty0 (3):\penalty0 4420--4437, 12 2021.
\newblock ISSN 0035-8711.
\newblock \doi{10.1093/mnras/staa3879}.
\newblock URL \url{https://doi.org/10.1093/mnras/staa3879}.

\bibitem[Oliver et~al.(2022)Oliver, Elahi, and Lewis]{Oliver2022}
William~H Oliver, Pascal~J Elahi, and Geraint~F Lewis.
\newblock {The hierarchical structure of galactic haloes: generalized N-dimensional clustering with CluSTAR-ND}.
\newblock \emph{Monthly Notices of the Royal Astronomical Society}, 514\penalty0 (4):\penalty0 5767--5785, 06 2022.
\newblock ISSN 0035-8711.
\newblock \doi{10.1093/mnras/stac1701}.
\newblock URL \url{https://doi.org/10.1093/mnras/stac1701}.

\bibitem[{Oliver} et~al.(2024){Oliver}, {Elahi}, {Lewis}, and {Buck}]{Oliver2024}
William~H. {Oliver}, Pascal~J. {Elahi}, Geraint~F. {Lewis}, and Tobias {Buck}.
\newblock {The hierarchical structure of galactic haloes: differentiating clusters from stochastic clumping with ASTROLINK}.
\newblock \emph{Monthly Notices of the Royal Astronomical Society}, 530\penalty0 (3):\penalty0 2637--2647, May 2024.
\newblock \doi{10.1093/mnras/stae1029}.

\bibitem[Onions et~al.(2012)Onions, Knebe, Pearce, Muldrew, Lux, Knollmann, Ascasibar, Behroozi, Elahi, Han, Maciejewski, Merchán, Neyrinck, Ruiz, Sgró, Springel, and Tweed]{Onions2012}
Julian Onions, Alexander Knebe, Frazer~R. Pearce, Stuart~I. Muldrew, Hanni Lux, Steffen~R. Knollmann, Yago Ascasibar, Peter Behroozi, Pascal Elahi, Jiaxin Han, Michal Maciejewski, Manuel~E. Merchán, Mark Neyrinck, Andrés~N. Ruiz, Mario~A. Sgró, Volker Springel, and Dylan Tweed.
\newblock {Subhaloes going Notts: the subhalo-finder comparison project}.
\newblock \emph{Monthly Notices of the Royal Astronomical Society}, 423\penalty0 (2):\penalty0 1200--1214, 06 2012.
\newblock ISSN 0035-8711.
\newblock \doi{10.1111/j.1365-2966.2012.20947.x}.
\newblock URL \url{https://doi.org/10.1111/j.1365-2966.2012.20947.x}.

\bibitem[Onions et~al.(2013)Onions, Ascasibar, Behroozi, Casado, Elahi, Han, Knebe, Lux, Merchán, Muldrew, Neyrinck, Old, Pearce, Potter, Ruiz, Sgró, Tweed, and Yue]{Onions2013}
Julian Onions, Yago Ascasibar, Peter Behroozi, Javier Casado, Pascal Elahi, Jiaxin Han, Alexander Knebe, Hanni Lux, Manuel~E. Merchán, Stuart~I. Muldrew, Mark Neyrinck, Lyndsay Old, Frazer~R. Pearce, Doug Potter, Andrés~N. Ruiz, Mario~A. Sgró, Dylan Tweed, and Thomas Yue.
\newblock {Subhaloes gone Notts: spin across subhaloes and finders}.
\newblock \emph{Monthly Notices of the Royal Astronomical Society}, 429\penalty0 (3):\penalty0 2739--2747, 01 2013.
\newblock ISSN 0035-8711.
\newblock \doi{10.1093/mnras/sts549}.
\newblock URL \url{https://doi.org/10.1093/mnras/sts549}.

\bibitem[{Planck Collaboration} et~al.(2014){Planck Collaboration}, {Ade}, {Aghanim}, {Armitage-Caplan}, {Arnaud}, {Ashdown}, {Atrio-Barandela}, {Aumont}, {Baccigalupi}, {Banday}, and et~al.]{Planck2014}
{Planck Collaboration}, P.~A.~R. {Ade}, N.~{Aghanim}, C.~{Armitage-Caplan}, M.~{Arnaud}, M.~{Ashdown}, F.~{Atrio-Barandela}, J.~{Aumont}, C.~{Baccigalupi}, A.~J. {Banday}, and et~al.
\newblock {Planck 2013 results. XVI. Cosmological parameters}.
\newblock \emph{Astronomy \& Astrophysics}, 571:\penalty0 A16, November 2014.
\newblock \doi{10.1051/0004-6361/201321591}.

\bibitem[Sain(2002)]{Sain2002}
Stephan~R. Sain.
\newblock Multivariate locally adaptive density estimation.
\newblock \emph{Computational Statistics \& Data Analysis}, 39\penalty0 (2):\penalty0 165--186, 2002.
\newblock ISSN 0167-9473.
\newblock \doi{https://doi.org/10.1016/S0167-9473(01)00053-6}.
\newblock URL \url{https://www.sciencedirect.com/science/article/pii/S0167947301000536}.

\bibitem[{Sestito} et~al.(2021){Sestito}, {Buck}, {Starkenburg}, {Martin}, {Navarro}, {Venn}, {Obreja}, {Jablonka}, and {Macci{\`o}}]{Sestito2021}
Federico {Sestito}, Tobias {Buck}, Else {Starkenburg}, Nicolas~F. {Martin}, Julio~F. {Navarro}, Kim~A. {Venn}, Aura {Obreja}, Pascale {Jablonka}, and Andrea~V. {Macci{\`o}}.
\newblock {Exploring the origin of low-metallicity stars in Milky-Way-like galaxies with the NIHAO-UHD simulations}.
\newblock \emph{Monthly Notices of the Royal Astronomical Society}, 500\penalty0 (3):\penalty0 3750--3762, January 2021.
\newblock \doi{10.1093/mnras/staa3479}.

\bibitem[Springel et~al.(2001)Springel, White, Tormen, and Kauffmann]{Springel2001}
Volker Springel, Simon D.~M. White, Giuseppe Tormen, and Guinevere Kauffmann.
\newblock Populating a cluster of galaxies – i. results at z = 0.
\newblock \emph{Monthly Notices of the Royal Astronomical Society}, 328\penalty0 (3):\penalty0 726--750, 2001.
\newblock ISSN 0035-8711.
\newblock \doi{10.1046/j.1365-8711.2001.04912.x}.
\newblock URL \url{https://doi.org/10.1046/j.1365-8711.2001.04912.x}.

\bibitem[Springel et~al.(2005)Springel, White, Jenkins, Frenk, Yoshida, Gao, Navarro, Thacker, Croton, Helly, Peacock, Cole, Thomas, Couchman, Evrard, Colberg, and Pearce]{Springel2005b}
Volker Springel, Simon D.~M. White, Adrian Jenkins, Carlos~S. Frenk, Naoki Yoshida, Liang Gao, Julio Navarro, Robert Thacker, Darren Croton, John Helly, John~A. Peacock, Shaun Cole, Peter Thomas, Hugh Couchman, August Evrard, Jörg Colberg, and Frazer Pearce.
\newblock Simulations of the formation, evolution and clustering of galaxies and quasars.
\newblock \emph{Nature}, 435\penalty0 (7042):\penalty0 629--636, 06 2005.
\newblock \doi{10.1038/nature03597}.

\bibitem[{Wang} et~al.(2015){Wang}, {Dutton}, {Stinson}, {Macci{\`o}}, {Penzo}, {Kang}, {Keller}, and {Wadsley}]{Wang2015}
L.~{Wang}, A.~A. {Dutton}, G.~S. {Stinson}, A.~V. {Macci{\`o}}, C.~{Penzo}, X.~{Kang}, B.~W. {Keller}, and J.~{Wadsley}.
\newblock {NIHAO project - I. Reproducing the inefficiency of galaxy formation across cosmic time with a large sample of cosmological hydrodynamical simulations}.
\newblock \emph{Monthly Notices of the Royal Astronomical Society}, 454:\penalty0 83--94, November 2015.
\newblock \doi{10.1093/mnras/stv1937}.

\end{thebibliography}

\end{document}